\documentclass[11pt]{article}
\pdfoutput=1
\usepackage{jheppub}

\usepackage[svgnames,table]{xcolor}
\usepackage{graphicx}
\usepackage{caption}
\usepackage{subcaption}
\usepackage{amsfonts,amsmath,amssymb,amsthm}
\usepackage{mathrsfs,bm,bbm,esint}
\usepackage[mathscr]{eucal}
\usepackage{physics}
\usepackage{dsfont}
\usepackage{slashed,soul}
\usepackage{rotating,pdflscape}
\usepackage{enumerate}
\usepackage{multirow,array}
\usepackage[numbers,sort&compress]{natbib}
\usepackage{tikz}
\usetikzlibrary{cd}
\usetikzlibrary{decorations.pathmorphing}
\usetikzlibrary{decorations.pathreplacing}
\usetikzlibrary{decorations.markings}
\tikzset{snake it/.style={decorate, decoration=snake}}
\tikzset{->-/.style={decoration={
  markings,
  mark=at position .5 with {\arrow{>}}},postaction={decorate}}}
\usepackage{subcaption}
\usepackage{float}
\usepackage{afterpage}
\renewcommand{\thefigure}{\arabic{figure}}

\def\be{\begin{eqnarray}}
\def\ee{\end{eqnarray}}

\def\t{\tilde}

\title{Saturation of Thermal Complexity of Purification}
 
 \author[a]{S. Shajidul Haque}
 \author[b]{, Chandan Jana}
\author[c]{, Bret Underwood}
\affiliation[a]{
High Energy Physics, Cosmology \& Astrophysics Theory Group and The Laboratory for Quantum Gravity \& Strings, Department of Mathematics and Applied Mathematics, \\ University of Cape Town, South Africa }
\affiliation[b]{Mandelstam Institute for Theoretical Physics, Witwatersrand University, Johannesburg, South Africa}
\affiliation[c]{Department of Physics, Pacific Lutheran University, Tacoma, WA 98447}

\emailAdd{shajid.haque@uct.ac.za }
\emailAdd{channdann.jana@gmail.com }
\emailAdd{bret.underwood@plu.edu}

\abstract{We purify the thermal density matrix of a free harmonic oscillator as a two-mode squeezed state, characterized by a squeezing parameter and squeezing angle. While the squeezing parameter is fixed by the temperature and frequency of the oscillator, the squeezing angle is otherwise undetermined, so that the complexity of purification is obtained by minimizing the complexity of the squeezed state over the squeezing angle. The resulting complexity of the thermal state is minimized at non-zero values of the squeezing angle and saturates to an order one number at high temperatures, indicating that there is no additional operator cost required to build thermal states beyond a certain temperature. We also review applications in which thermal density matrices arise for quantum fields on curved spacetimes, including Hawking radiation and a simple model of decoherence of cosmological density perturbations in the early Universe. The complexity of purification for these mixed states also saturates as a function of the effective temperature, which may have interesting consequences for the quantum information stored in these systems.
}

\begin{document}
\maketitle
\raggedbottom

\section{Introduction}
\label{sec:intro}

The complexity of a quantum circuit, defined as the minimum number of unitary operations that are needed to transform a given reference state into a particular target state \cite{NL1,NL2,NL3,Jefferson}, has many interesting applications.
There appear to be interesting connections to gravitational holography, where the complexity of a field theory living on a boundary may characterize features of its gravitational dual \cite{Susskind:2014rva,Stanford:2014jda,Brown:2015bva}.
Quantum circuit complexity (hereafter simply referred to as complexity) may also serve as a diagnostic for quantum chaos \cite{Ali:2019zcj,Bhattacharyya:2019txx,Bhattacharyya:2020art}, 
and the complexity of quantum cosmological perturbations shows interesting behaviors \cite{cosmology1,cosmology2,Lehners_2021}.

While pure states are simple to work with, many interesting systems and phenomena are described by mixed states,
including decoherence and thermal states.
A mixed state can be transformed into a pure state through purification, in which the Hilbert space is enlarged to include ancillary degrees of freedom such that the mixed state density matrix is recovered as the trace of the pure state density matrix over the ancillary states.
We will follow \cite{Agn2019,MyersMixed,DiGiulio:2020hlz,Camargo:2020yfv} in defining the \emph{thermal complexity of purification} as the minimum complexity of a set of purifications of a thermal density matrix relative to the ground state.
Alternatively, a pure state can be associated to the density matrix through the technique of operator-state mapping \cite{CHOI1975285,JAMIOLKOWSKI1972275}, in such a way that the expectation values of observables are preserved.
We will consider both techniques in Section \ref{sec:sp_purifcn}, finding that the extra freedom in scanning over ancillary states for the thermal complexity of purification leads to a smaller complexity (and thus a more optimal unitary operator) than the unique pure state defined by the operator-state mapping procedure.

We will focus on the thermal density matrix of a free harmonic oscillator, both for its simplicity and because it can be directly related to several interesting applications. 
While the thermal complexity of purification for this system has been studied before \cite{MyersMixed,DiGiulio:2020hlz,Camargo:2020yfv} (see also \cite{Bhattacharyya:2018sbw,Bhattacharyya:2019tsi} for the entanglement of purification of similar systems), 
our analysis, based on the approach of \cite{Jefferson,me1}, will include a somewhat more general purification as a two-mode squeezed state with non-zero squeezing angle $\phi$.
Squeezed states can play an important role in continuous variable quantum computing \cite{TeleportQuantum,Furusawa706,Braunstein_2005,qumode,Fukui_2018}, and show up naturally in descriptions of Hawking radiation and cosmological perturbations \cite{Grishchuk,Albrecht,Martin1,Martin2}, among other applications.
We will find in Section \ref{sec:sub:TwoModePurif} that the minimum thermal complexity of purification relative to the ground state occurs at a non-zero squeezing angle, and saturates at high temperature $T$ (or low frequencies $\omega$) to a constant ${\mathcal C}_{\rm th} \approx \pi/(2\sqrt{2})$, as opposed to the complexity for a vanishing squeezing angle that grows logarithmically with temperature ${\mathcal C}_{\phi = 0} \sim \ln(T/\omega)$.
In Section \ref{sec:sub:AdditionalSqueeze}, we show that generalizing the purification further to also include single-mode squeezing of the ancillary degrees of freedom introduces additional degeneracy in the purification parameters for the minimized complexity, but does not change the value of the minimized complexity.
In Section \ref{sec:sub:OperatorState}, we show how the pure state obtained from operator-state mapping is a two-mode squeezed state with a vanishing squeezing angle, and thus does not have a minimal complexity. Altogether, we find that the two-mode squeezed complexity of purification with a non-zero squeezing angle produces the simplest minimal complexity among these different purification techniques.

We apply our results on the thermal complexity of purification to two interesting applications in quantum fields on curved spacetime where thermal density matrices can arise: Hawking radiation and cosmological perturbation theory.
In Section \ref{sec:sub:Hawking}, we review how Hawking radiation for a scalar field on a curved spacetime with horizons, such as Rindler space or a black hole spacetime, can be described as a thermal state by tracing over a two-mode squeezed state in which modes on either side of the horizon are entangled with each other.
We then illustrate how the optimal purification of complexity of this thermal state takes the form of a squeezed state with a different squeezing angle and saturates at when the effective temperature of the horizon is high, in contrast to the complexity of the original squeezed state which grows with the effective temperature.
While these are calculations of complexity ``in the bulk'' (and are thus not directly related to holography), they may have interesting implications for information-theoretic perspectives of black holes.
In Section \ref{sec:sub:SqueezedCosmo}, we review the application of the techniques of complexity to quantum cosmological perturbations, such as those produced in the very early universe during inflation.
Considering a few simple models of decoherence, we show how the Fourier modes of these perturbations can also be described by a thermal density matrix, and illustrate how the resulting saturation of the thermal complexity of purification of the Universe compares before and after decoherence.
We conclude in Section \ref{sec:Discussion} with a summary of our results and some speculation on their implications.
Several appendices are included for reference and review.


\section{Complexity of a Thermal Density Matrix}
\label{sec:sp_purifcn}

As discussed in the Introduction, we are interested in the complexity of a thermal state of a quantum harmonic oscillator,
\be
\hat\rho_{\rm th} = \frac{1}{Z} \sum_{n=0}^\infty e^{-\beta E_n} |n\rangle\langle n|\, ,
\label{eq:thermalRho}
\ee
where $E_n = \omega n$.
In order to use established techniques \cite{Jefferson,me1} to compute the complexity, we need to represent the thermal state (\ref{eq:thermalRho}) as a pure state $|\Psi\rangle$. 
%
%
%
%
For any mixed state $\hat \rho_{\rm mix}$ on the Hilbert space $\cal H$, we can construct a \emph{purification} of $\hat \rho_{\rm mix}$ which consists of a pure state $|\Psi\rangle$ in an enlarged Hilbert space ${\cal H}_{\rm pure} = \cal H \otimes \cal H_{\rm anc}$, where ${\cal H}_{\rm anc}$ corresponds to an ``ancillary'' set of degrees of freedom. 
If the trace of the density matrix of $|\Psi\rangle$ over the ancillary degrees of freedom gives the original mixed state 
${\rm Tr}_{\rm anc}\left(|\Psi\rangle\langle \Psi|\right) = \hat \rho_{\rm mix}$,
we say that $|\Psi\rangle$ is a ``purification'' of $\hat \rho_{\rm mix}$.
Note that expectation values of operators acting in ${\cal H}$ are preserved under purification, $\langle \hat {\cal O}\rangle = {\rm Tr}_{\rm anc}\left(\langle \Psi|\hat {\cal O}|\Psi\rangle\right) = {\rm Tr}\left(\hat \rho_{\rm mix}\hat {\cal O}\right)$, so that observables are preserved by purification.

Clearly, the purification $|\Psi\rangle$ is not unique, since the choice of the ancillary Hilbert space ${\cal H}_{\rm anc}$ is arbitrary as long as it meets the purification requirement.
For example, there may be a set of pure states $\left\{|\Psi\rangle_{\alpha,\beta,...}\right\}$, parameterized by $\alpha, \beta, ...$, all of which satisfy the purification requirement.
In order to distinguish among the set of purifications, it is often helpful to minimize a quantity of interest, such as the entanglement entropy or complexity, with respect to the parameters.
In this work, we are interested in analyzing the complexity of the mixed thermal state (\ref{eq:thermalRho}), so we will minimize the complexity of the set of purifications $\left\{|\Psi\rangle_{\alpha,\beta,...}\right\}$ of $\hat \rho_{\rm th}$, obtaining the {\bf thermal complexity of purification}
\be
{\cal C}_{\rm th}(\beta) = \min_{\alpha, \beta,...} {\cal C}\left(|\Psi\rangle_{\alpha, \beta,...},|\psi_R\rangle\right)\, ,
\label{eq:CoP}
\ee
where we made explicit the dependence of the complexity of the pure state on the reference state $|\psi_R\rangle$.

We will choose three explicit purifications of $\hat \rho_{\rm th}$ (\ref{eq:thermalRho}).
First, in Section \ref{sec:sub:TwoModePurif} we will purify $\rho_{\rm th}$ as a two-mode squeezed state, parameterized by a squeezing parameter $r$ and squeezing angle $\phi$, in which the Hilbert spaces ${\cal H},{\cal H}_{\rm anc}$ are entangled in such a way that a trace over the ancillary degree of freedom gives rise to a thermal state. The corresponding complexity of purification \emph{saturates} as a function of inverse temperature.
Next, in Section \ref{sec:sub:AdditionalSqueeze} we generalize the two-mode squeezed state purification to include additional squeezing of the ancillary degree of freedom.
We find that minimizing over the additional parameters introduced by the extra squeezing provides more freedom in finding a minimum of (\ref{eq:CoP}), but leaves the minimum value of the complexity of purification unchanged.
Finally, in Section \ref{sec:sub:OperatorState} we construct a purification of the thermal state through operator-state mapping.

\subsection{Two Mode Squeezing as Purification}
\label{sec:sub:TwoModePurif}

A straightforward purification of the generic thermal state (\ref{eq:thermalRho}) is the thermofield double state,
\be
|{\rm TFD}\rangle = \frac{1}{\sqrt{Z}} \sum_{n=0}^\infty e^{-\beta E_n/2} |n\rangle \otimes |n\rangle_{\rm anc}\, ,
\label{eq:TFD}
\ee
where the ancillary Hilbert space ${\cal H}_{\rm anc}$ is taken to be a copy of the original oscillator ${\cal H}$.
However, the thermofield double state (\ref{eq:TFD}) is by no means unique as a purification of (\ref{eq:thermalRho}); 
indeed, it is possible to include an additional phase, so that the purification $|\Psi\rangle_{\alpha, \beta...}$ becomes
\be
|\Psi\rangle_{\phi} = \vert {\rm TFD} \rangle_\phi = \frac{1}{\sqrt{Z}} \sum_{n=0}^\infty (-1)^{n} e^{-2in\phi} e^{-n \beta\omega/2} | n \rangle \otimes | n \rangle_{\rm anc} \, ,
\label{eq:TFDphi}
\ee
where we took $E_n = \omega n$ for bosonic oscillators, and we introduced a factor of $(-1)^n$ (which can be reabsorbed back into $\phi$) for convenience.
We recognize this as a two-mode squeezed vacuum state,
\be
|\Psi\rangle_{\phi} = \frac{1}{\cosh r} \sum_{n=0}^\infty (-1)^{n} e^{-2i n\phi} \tanh^n r | n \rangle \otimes | n \rangle_{\rm anc} \equiv \hat S_{\rm sq}(r,\phi) | 0\rangle \otimes |0 \rangle_{\rm anc}\, ,
\label{eq:TFDsqueeze}
\ee
where the squeezing parameter $r$ is related to the oscillator frequency and temperature through $\beta\omega = - \ln \tanh^2 r$, and the squeezing angle $\phi$ is a free parameter; as part of the purification of complexity process (\ref{eq:CoP}), we will minimize the complexity with respect to $\phi$.
The operator $\hat S_{\rm sq}(r,\phi)$ is the two-mode squeezing operator, and is given in terms of raising and lowering operators $\{\hat a, \hat a^\dagger\},\{\hat a_{\rm anc},\hat a_{\rm anc}^\dagger\}$ on the physical and ancillary oscillator Hilbert spaces ${\cal H},{\cal H}_{\rm anc}$, respectively,
\begin{equation}
	\hat S_{\rm sq}(r,\phi) = \exp \left[ \frac{r}{2} \left( e^{-2i\phi} \hat a \hat a_{\rm anc}  - e^{2i\phi} \hat a^\dagger \hat a_{\rm anc}^\dagger \right) \right] \, .
\end{equation}
The purification of the thermal state (\ref{eq:thermalRho}) into (\ref{eq:TFDsqueeze}) is thus obtained by acting the squeeze operator $\hat S_{\rm sq}(r,\phi)$ on the two-mode vacuum $|0\rangle \otimes|0\rangle_{\rm anc}$. 
The two-mode squeezing operator can be interpreted as a type of \emph{entanglement} operator, as it mixes creation and annihilation operators of
the two Hilbert spaces ${\cal H},{\cal H}_{\rm anc}$
\be
\hat S^\dagger_{\rm sq}\ \hat a\ \hat S_{\rm sq} &=& \hat a\ \cosh r - \hat a_{\rm anc}^\dagger\ e^{2i\phi}\ \sinh r\, ; \\
\hat S^\dagger_{\rm sq}\ \hat a_{\rm anc}\ \hat S_{\rm sq} &=& \hat a_{\rm anc}\ \cosh r - \hat a^\dagger\ e^{2i\phi}\ \sinh r\, .
\ee

For the calculation of complexity\footnote{See Appendix \ref{app:Complexity} for details.}, 
we are interested in the transformation of a reference state
$|\psi_R\rangle$ into a target state $|\psi_T\rangle$
\be
|\psi_T\rangle = \hat {\mathcal U}\ |\psi_R\rangle
\label{eq:TargetReference}
\ee
by a unitary operator $\hat {\mathcal U}$ representing the quantum circuit connecting these two states.
Following the geometric approach of \cite{NL1,NL2,NL3,Jefferson}, 
we decompose the unitary $\hat {\mathcal U}$ as a path-ordered sequence of a set of fundamental operators $\{\hat {\mathcal O}_I\}$
\be
\hat U(s) = \overleftarrow{\mathcal P}\ {\rm exp}\left[-i \int_0^s \sum_I Y^I(s') \hat {\mathcal O}_I\  ds'\right]\, .
\label{eq:Unitary}
\ee
We define complexity in a geometric way as the circuit depth along a minimal path in the geometry generated by the algebra of the operators
\be
{\mathcal C} = \int_0^1 \sqrt{\sum_I G_{IJ} Y^I Y^J}\ ds\, ,
\label{eq:ComplexityNielsen}
\ee
where the $Y^I(s)$ are vectors that specify the path parameterized by $s$.
To calculate the complexity 
we follow \cite{Jefferson,cosmology1} in transforming 
our reference and target states into position-space wavefunctions
by defining the position variable $\hat q = \frac{1}{\sqrt{2\omega}} (\hat a^\dagger + \hat a)$ (with a similar definition for the ancillary variable $\hat q_{\rm anc}$ in terms of the $\hat a_{\rm anc},\hat a_{\rm anc}^\dagger$).
Our reference state will be the (purified) ground state $|\psi_R\rangle = |0\rangle \otimes|0\rangle_{\rm anc}$
\be
\langle q,q_{\rm anc}|\psi_R\rangle = {\mathcal N}_R\ {\rm exp}\left[-\frac{1}{2} \omega (q^2 + q_{\rm anc}^2)\right]\, .
\label{eq:gaussianRef}
\ee
Our target state is the purification (\ref{eq:TFDsqueeze}), $|\psi_T\rangle = |\Psi\rangle_{\phi}$, which also takes the form of a Gaussian wavefunction
\be
\Psi_{\rm sq}\left(q,q_{\rm anc}\right) = \langle q, q_{\rm anc}|\Psi\rangle_{\phi} = {\cal N} \exp\left\{-\frac{\omega}{2}\, A (q^2 + {q}_{\rm anc}^2) - \omega B\ q\ q_{\rm anc} \right\}
\label{eq:psiSqueezed}
\ee
where ${\cal N}$ is a normalization factor that will not be important here, and the squeezed Gaussian parameters are
\begin{equation}
	\begin{split}
		&
		A = \frac{1 + e^{-4i\phi} \tanh^2 r}{1-e^{-4i\phi} \tanh^2 r} \,,  
		\qquad
		B = \frac{2\tanh r\, e^{-2i\phi}}{1- e^{-4i\phi} \tanh^2 r} \,.
	\end{split}\label{eq:gaussianAB}
\end{equation}
Interestingly, the corresponding variances of the physical position $\hat q$ and momentum $\hat p = i\sqrt{\omega/2} (\hat a^\dagger - \hat a)$ are independent of the squeezing angle
\be
 \langle q^2 \rangle = \frac{1}{\omega}\cosh(2r)\, ,\qquad \langle p^2\rangle = \omega \cosh(2r)\, .
\ee
A curious feature of the two-mode squeezed state is that the uncertainty of the single oscillator grows as the squeezing is increased $\langle q^2\rangle \langle p^2\rangle = \cosh^2(2r)$, in contrast to a one-mode squeezed state in which the uncertainty in one direction of phase space is reduced while the other grows, preserving the total uncertainty\footnote{Two mode squeezed states do not preserve the total noise, but instead preserve the difference in the total noises of the two modes. See \cite{Schumaker} for other useful properties of two-mode squeezed states.}.
The growth in uncertainty of the single oscillator of the two-mode squeezed state mirrors the expected growth in uncertainty of a thermal state with temperature, and is a desired property of our purification of the thermal mixed state.


The complexity (\ref{eq:ComplexityNielsen}) of the purified thermal state (\ref{eq:psiSqueezed}) thus becomes (see Appendix \ref{app:Complexity})
\begin{eqnarray}\label{eqn:squeeze_vac_complexity}
	\mathcal{C}_{\phi} &=& \frac{1}{\sqrt{2}} \sqrt{\ln^2\left| \frac{1 +e^{-2i\phi} \tanh r}{1 -e^{-2i\phi} \tanh r} \right| + \arctan^2(\sin 2\phi \sinh 2r)} \\
	&=& \frac{1}{\sqrt{2}} \sqrt{\ln^2\left| \frac{1 +e^{-2i\phi} e^{-\beta \omega/2}}{1 -e^{-2i\phi} e^{-\beta \omega/2}} \right| +  \arctan^2\left(2\sin 2\phi \frac{e^{-\beta\omega/2}}{1-e^{-\beta\omega}}\right)} \,,
	\label{eqn:squeeze_vac_complexity2}
\end{eqnarray}
where we substituted $\tanh r = e^{-\beta\omega/2}$ to make the temperature-dependence explicit.
The $\arctan$ contribution to the complexity is necessary when the parameters $A,B$ of the Gaussian wavefunction (\ref{eq:psiSqueezed}) take on complex values \cite{me1}, and will play an important role in the minimized complexity, as we will see.

\begin{figure}[t]
		\centering
		\includegraphics[width=.6\textwidth]{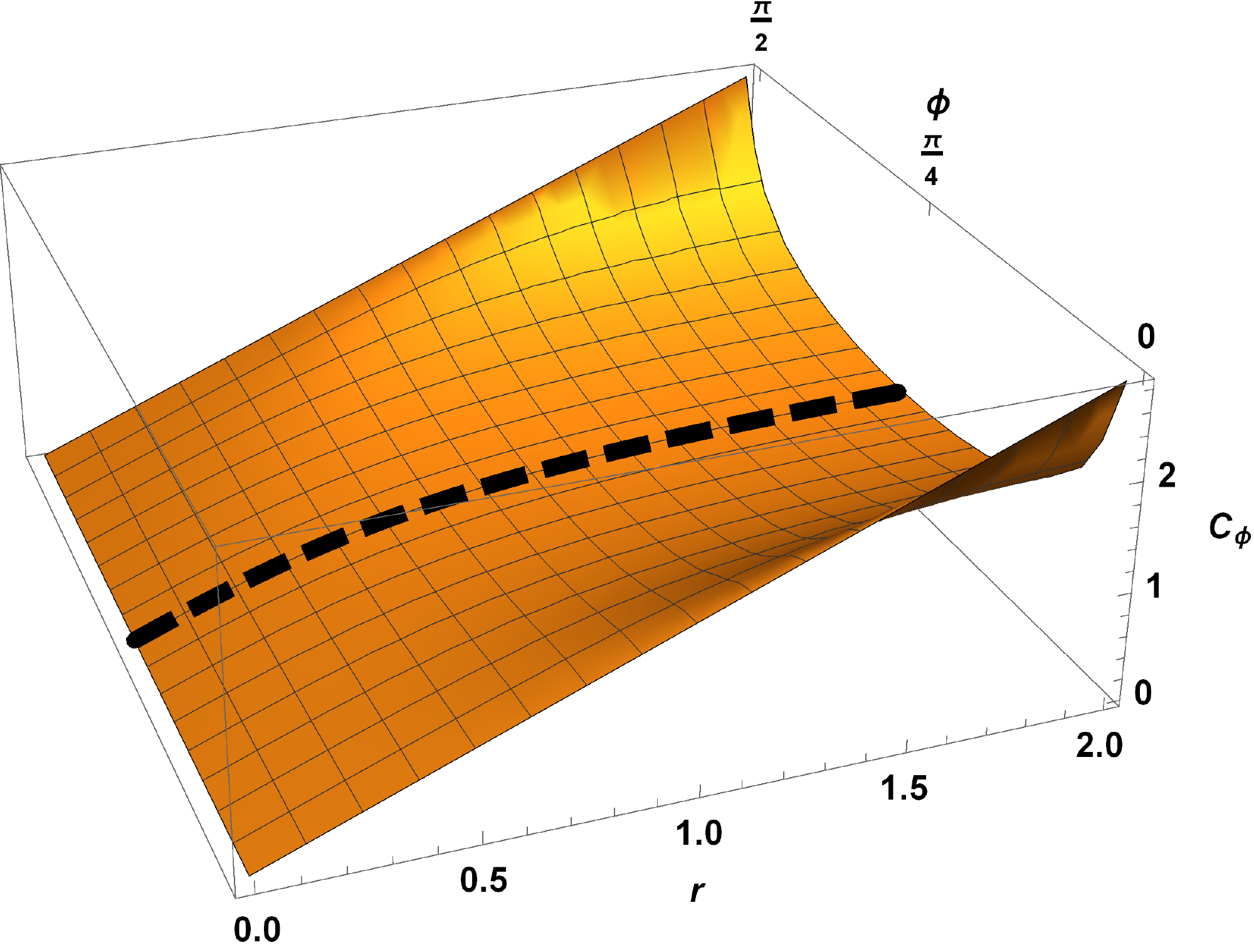}
		\caption{Complexity (\ref{eqn:squeeze_vac_complexity}) of the purified thermal state as a squeezed state, as a function of the squeezing parameter $r$ and the squeezing angle $\phi$. The dashed line indicates the minimum of the complexity at $\phi = \pi/4$.
			The complexity saturates at large $r$ (corresponding to high temperature) for generic $\phi$ as in (\ref{eq:squeezeSaturateGeneral}), but grows linearly with the squeezing $r$ for the angles $\phi = 0, \pi/2$.}
		\label{fig:Cphi3D2}
\end{figure}

Before we minimize the complexity (\ref{eqn:squeeze_vac_complexity2}) with respect to the squeezing angle $\phi$ as part of the complexity of purification process (\ref{eq:CoP}), let us examine some of its useful limits.
The low-temperature limit $\beta \omega \rightarrow \infty$ corresponds to vanishing squeezing $r \rightarrow 0$; 
the system and ancillary degrees of freedom are no longer entangled, as can be seen by the diagonalization of the wavefunction (\ref{eq:psiSqueezed}) for $A \approx 1, B \approx 0$ in this limit.
Correspondingly, the complexity (\ref{eqn:squeeze_vac_complexity2}) vanishes as ${\cal C}_{\phi} \sim e^{-\beta \omega/2}$ at low temperatures $\beta \omega \gg 1$, for all squeezing angles $\phi$.

Alternatively, the high-temperature limit $\beta \omega \ll 1$ corresponds to large squeezing $r \gg 1$ in the squeezed-state language.
For generic squeezing angles (specifically, for $\phi \neq n\pi/2$) the Gaussian wavefunction parameters (\ref{eq:gaussianAB}) saturate to pure imaginary values at leading order
\be
A \approx -i \frac{\sin 4\phi}{1-\cos 4\phi}\, , \qquad B \approx -2 i \frac{\sin 2\phi}{1-\cos 4\phi}\, .
\label{eq:ABImaginary}
\ee
Thus, at high temperature (large squeezing), the two-mode squeezed state wavefunction (\ref{eq:psiSqueezed}) is an approximately pure phase entanglement between the physical and ancillary degrees of freedom.
Correspondingly, the position variance diverges in the high-temperature limit $\omega \langle q^2 \rangle \sim (\beta\omega)^{-1} \gg 1$, consistent with the delocalization that is expected of a thermal state.
The saturation of the purified wavefunction to a pure phase at high temperatures leads to a similar saturation of the complexity (\ref{eqn:squeeze_vac_complexity2})
\be
{\mathcal C}_\phi \approx \frac{1}{\sqrt{2}} \sqrt{\ln^2 \left|\frac{1+\cos2\phi}{1-\cos2\phi}\right| + \left(\frac{\pi}{2}\right)^2}\, ,
\label{eq:squeezeSaturateGeneral}
\ee
which is independent of the temperature at leading order.
The angle $\phi = \pi/4$ is particularly interesting, since we see that the entangled Gaussian wavefunction becomes purely off-diagonal in the high-temperature limit $A\approx 0, B \approx i$, and the complexity is minimal ${\mathcal C}_{\phi/4} \approx \pi/2\sqrt{2}$.

The angles $\phi = n\pi/2$ are special cases, and lead to behaviors for the wavefunction and complexity that are qualitatively different
than the general case.
For example, for a vanishing squeezing angle $\phi = 0$, the high-temperature limit corresponds to a strongly entangled wavefunction
with real Gaussian parameters $A \sim B\sim (\beta \omega)^{-1} \gg 1$.
The corresponding complexity (\ref{eqn:squeeze_vac_complexity2}) is then dominated by the first term ${\cal C}_{\phi = 0} \approx r \approx \ln\left(\frac{1}{\beta\omega}\right) = \ln \left(\frac{T}{\omega}\right)$ and grows logarithmically with the temperature. 
The generic behavior of the complexity, as well as the behavior for the special cases $\phi = n\pi/2$, 
can be seen in Figure \ref{fig:Cphi3D2}.

\begin{figure}[t]
	\centering
	\begin{subfigure}[b]{0.49\textwidth}
		\centering
		\includegraphics[width=\textwidth]{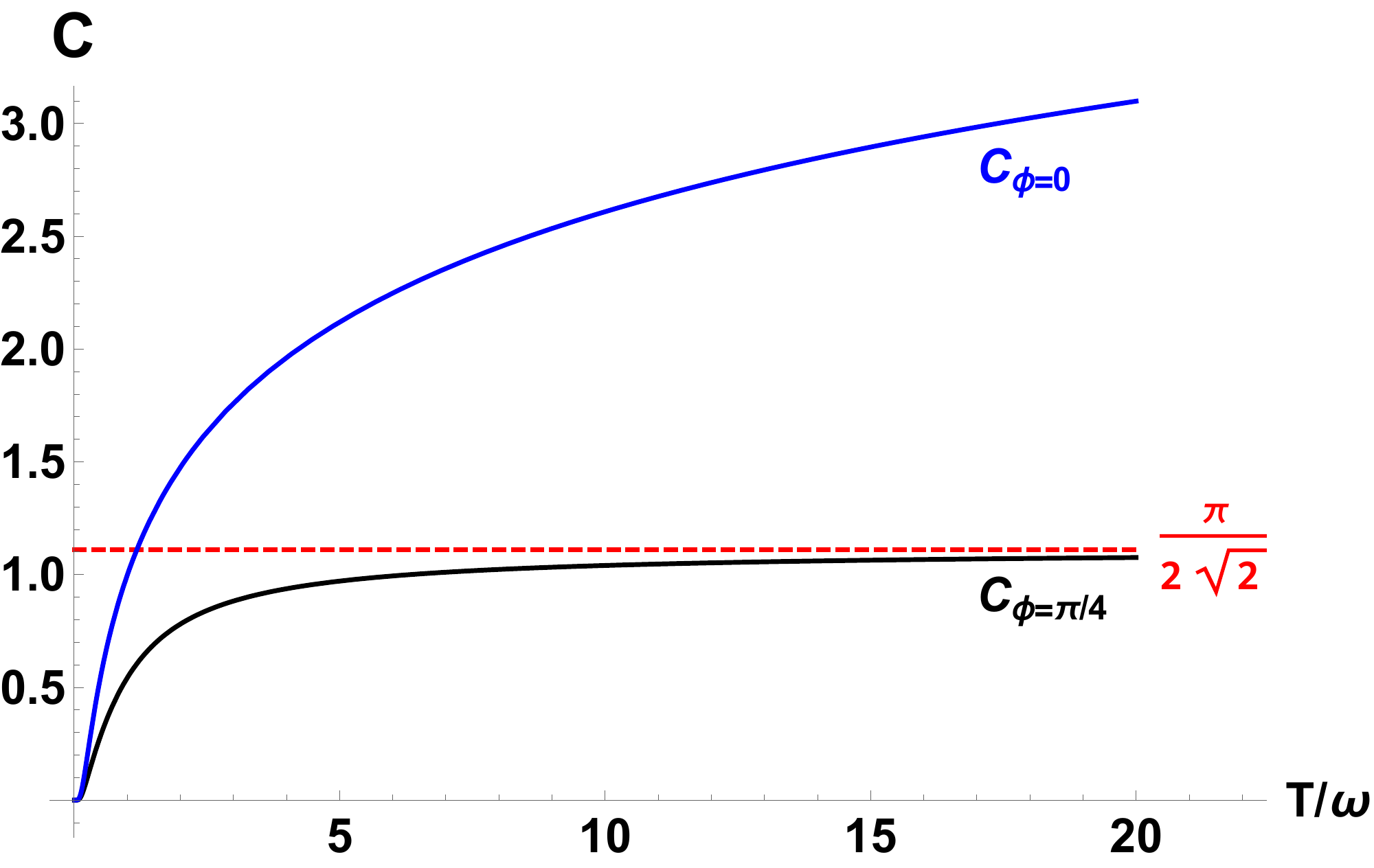}
		\caption{Complexity of purification (\ref{eqn:squeeze_vac_complexity2}) as a function of temperature $T = \beta^{-1}$ for $\phi = 0$ and $\phi = \pi/4$.}
		\label{subfig:Cp}
	\end{subfigure}\hfill
	\begin{subfigure}[b]{0.49\textwidth}
		\centering
		\includegraphics[width=\textwidth]{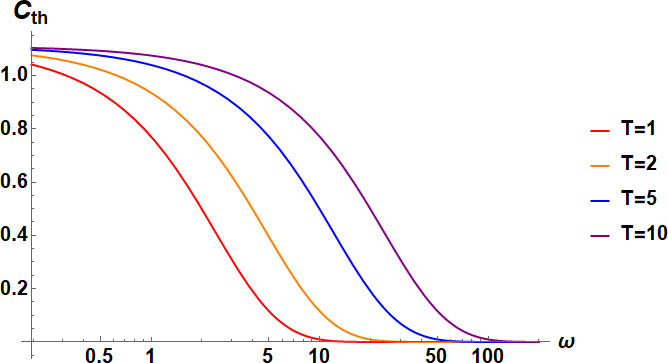}
		\caption{Thermal complexity of purification as a function of the frequency $\omega$ for different fixed temperatures $T$.}
		\label{subfig:Cp_omega}
	\end{subfigure}
	\caption{The complexity of purification is minimized by the squeezing angle $\phi = \pi/4$ and has the unique feature that it saturates for high temperatures to ${\mathcal C}_{\phi = \pi/4} \approx \pi/(2\sqrt{2}) \approx 1.1$, as seen in Figure \ref{subfig:Cp}. For a fixed temperature $T$, the minimal thermal complexity of purification ${\mathcal C}_{\rm th} = {\mathcal C}_{\phi = \pi/4}$ drops off sharply for high frequencies $\omega > T$, as seen in Figure \ref{subfig:Cp_omega}.}
	\label{fig:Cp}
\end{figure}

We have purified the thermal state $\hat \rho_{\rm th}$ (\ref{eq:thermalRho}) into the two-mode squeezed state (\ref{eq:TFDsqueeze}), with corresponding complexity (\ref{eqn:squeeze_vac_complexity2}). 
The {\bf thermal complexity of purification} (\ref{eq:CoP}) is obtained by minimizing (\ref{eqn:squeeze_vac_complexity2}) over the squeezing angle $\phi$, which we introduced as a free parameter.
The squeezed state complexity (\ref{eqn:squeeze_vac_complexity2}) is symmetric about $\pi/4$, and the complexity is minimized at $\phi = \pi/4$ for all values of the squeezing parameter $r$ (or correspondingly, for all values of the temperature $\beta \omega$) leading to
\be
	\label{eq:Cp}
	\mathcal{C}_{\rm th}(\beta) &=& \mathcal{C}_\phi\big\vert_{\phi=\pi/4} = \frac{1}{\sqrt{2}} |\arctan(\sinh 2r)| = \frac{1}{\sqrt{2}}\left|\arctan \left(2\frac{e^{-\beta \omega/2}}{1-e^{-\beta \omega}}\right)\right|   \\
	&\approx& \begin{cases}\sqrt{2}\ e^{-\beta \omega/2} & \mbox{ low temperature limit } \beta \omega \gg 1 \cr \frac{\pi}{2\sqrt{2}} & \mbox{ high temperature limit, } \beta \omega \rightarrow 0\end{cases}
\ee
with the corresponding purification as a two-mode squeezed state with squeezing angle $\phi = \pi/4$
\be
|\Psi\rangle_{\rm th,p} = |\Psi\rangle_{\phi = \pi/4} = \frac{1}{\sqrt{Z}}\sum_{n=0}^\infty (-i)^n\ e^{-n\beta\omega/2} |n\rangle \otimes |n\rangle_{\rm anc}\, .
\label{eq:PsiPurif}
\ee
Interestingly, we find that the complexity of purification of the thermal density matrix $\hat \rho_{\rm th}$ \emph{saturates} at high temperature, as can also be seen in Figure \ref{fig:Cp}.
This saturation of the complexity of the purified thermal density matrix with temperature is in contrast to previous results \cite{MyersMixed}, which did not include the squeezing angle in the process of purification, and thus considered only real Gaussian wavefunctions.
We see that purifying the thermal density matrix to include a squeezing angle qualitatively changes the behavior of the complexity of purification.
As discussed above, the physical origin of this saturation is that at high temperature the Gaussian wavefunction parameters (\ref{eq:ABImaginary}) become imaginary and independent of the temperature at leading order.
The corresponding circuit depth saturates, even for arbitrarily large temperature.

It is interesting to compare the thermal complexity of purification to the corresponding entanglement entropy of the thermal state \cite{Srednicki_1993,demarie2012pedagogical}
\be
\hat S_{\rm en} = {\rm Tr}\left[\hat \rho_{\rm th}\ln \hat \rho_{\rm th}\right] = -\ln\left(1-e^{-\beta \omega}\right) - \frac{e^{-\beta \omega}}{1-e^{-\beta \omega}}\ln \left(e^{-\beta \omega}\right)\, .
\label{eq:EEsqueezed}
\ee
The entanglement entropy of the thermal state is independent of the purification squeezing angle $\phi$, so that
from the perspective of entanglement entropy, all two-mode squeezed states that comprise the purification state (\ref{eq:TFDsqueeze})
are equivalent, for any squeezing angle.
However, we have seen that complexity for building such states is not equivalent -- it is in fact easier, by a factor $\sim |\ln(\beta \omega)|$, to build the optimal purification state (\ref{eq:PsiPurif}) ((\ref{eq:TFDsqueeze}) with $\phi = \pi/4$) than it is to build the TFD state (\ref{eq:TFD}) ((\ref{eq:TFDsqueeze}) with $\phi = 0$), even though both of these states give identical reduced thermal density matrices and have identical entanglement entropies (\ref{eq:EEsqueezed}).
This is in line with other observations that complexity is a more sensitive probe of a state than entanglement alone \cite{Camargo_2019,MyersMixed}.

Another interesting perspective of the thermal complexity of purification (\ref{eq:Cp}) is to consider its dependence on the oscillator frequency $\omega$ for fixed temperature $\beta = 1/T$, as in Figure \ref{subfig:Cp_omega}.
For fixed temperature, the complexity drops off quickly as a function of frequency ${\mathcal C}_{\rm th} \sim e^{-\beta\omega/2}$, so that high frequency modes with $\omega > T$ contribute less to the complexity than low frequency modes.
The suppression of the complexity for purified high frequencies of a thermal state characterizes the extent to which these modes are washed out by the thermal background.

The thermal state of a single harmonic oscillator is easily generalizable to a set of $N$ decoupled harmonic oscillators with Hamiltonian $\hat H = \sum_{i=1}^N \left(\omega_i \hat a_i^\dagger \hat a_i+1/2\right)$, each of which is in a thermal mixed state with individual thermal density matrix given by $\rho_{\rm th}$ (\ref{eq:thermalRho}).
The two-mode purification described above simply becomes a direct product of the individual purifications, so that the total thermal complexity of purification for all the modes is the sum
\be
{\mathcal C}_{\rm th}^{\rm N} = \frac{1}{\sqrt{2}} \sqrt{\sum_{i=1}^N \arctan^2\left(2\frac{e^{-\beta \omega_i/2}}{1-e^{-\beta\omega_i}}\right)} \approx \frac{1}{\sqrt{2}} \sqrt{\sum_{\omega_i < T} \arctan^2\left(2\frac{e^{-\beta \omega_i/2}}{1-e^{-\beta\omega_i}}\right)}\, ,
\ee
where only the low frequencies $\omega_i < T$ contribute to the sum since the $\arctan$ drops off at high frequencies.
Unfortunately, while the ${\mathcal F}_2$ cost function used here for the complexity (see Appendix \ref{app:Complexity}) has the advantage of being independent of basis choice, it is clearly not extensive with respect to the number of independent oscillators.
Instead, it is more natural to use the Finsler ${\mathcal F}_1$ cost function to include this behavior.
Fortunately the minimum length geodesic is unchanged, so that the (Finsler) thermal complexity of purification, minimized as a function of the squeezing angle $\phi$, becomes
\be
\label{eq:FinslerComplexity}
{\mathcal C}_{1}^{\rm N} &=& \sum_{i=1}^N \arctan\left(2\frac{e^{-\beta \omega_i/2}}{1-e^{-\beta\omega_i}}\right) \approx \sum_{\omega_i < T} \arctan\left(2\frac{e^{-\beta \omega_i/2}}{1-e^{-\beta\omega_i}}\right) \\
&\approx& \frac{\pi}{2\sqrt{2}}\ N_T\, ,
\ee
where $N_T$ is the number of oscillators with $\omega_i < T$.

It is straightforward to now take the continuum limit of (\ref{eq:FinslerComplexity}) as the complexity of the purification of Fourier modes of a free  scalar field $\varphi(x,t)$ in $(d+1)$-spacetime dimensions at constant temperature
\be
{\mathcal C}_{\rm th}^{\rm tot} = {\rm Vol}_d \int d^dk \arctan\left(2\frac{e^{-\beta \omega_{\vec{k}}/2}}{1-e^{-\beta \omega_{\vec{k}}}}\right) 
= {\rm Vol}_d\ {\cal I}_d\ \beta^{-d} \, ,
\ee
where ${\rm Vol}_d$ is the $d$-dimensional spatial volume, $\omega_{\vec{k}} = \sqrt{\vec{k}^2 + m^2}$ and 
\be
{\cal I}_d = \int d^d\ell \arctan\left(2\frac{e^{-\sqrt{\ell^2+m^2\beta^2}/2}}{1-e^{-\sqrt{\ell^2+m^2\beta^2}}}\right)
\ee
is a numerical constant.
For example, for $m= 0$ and $d = 1$, ${\cal I}_1 \approx 3.7$, the thermal complexity of purification grows linearly with the temperature, 
while for $m=0$ and $d = 3$, ${\cal I}_3 \approx 398$, the thermal complexity grows as the cube of the temperature.
The temperature dependence arises due to the density of states, since the complexity saturates for each mode with frequency $k < T$ in a sphere with radius set by the temperature.

We have considered the complexity of the purified target state (\ref{eq:PsiPurif}) relative to the ground state, but another natural candidate for a reference state is the factorized Gaussian state \cite{Jefferson,MyersMixed}
\be
\langle q,q_{\rm anc}|\psi_f\rangle = {\mathcal N}_{f} \exp\left[-\frac{1}{2} \mu (q^2 + q_{\rm anc}^2)\right]\, ,
\ee
where the reference frequency $\mu$ is the same for all oscillators and is not equal to the ground state frequency $\omega$. 
For a set of oscillators, this reference state is disentangled in the position basis, so the complexity of the target state (\ref{eq:PsiPurif}) relative to it is therefore a useful measure of the difficulty in creating spatial entanglement.
Minimizing the purified complexity with respect to the squeezing angle, we now have an additional term in the thermal complexity of purification
\be
{\mathcal C}_{\rm th,\mu}(\beta) = \frac{1}{\sqrt{2}} \sqrt{\left(\ln\left|\frac{\omega}{\mu}\right|\right)^2 + \arctan^2\left(2 \frac{e^{-\beta\omega/2}}{1-e^{-\beta\omega}}\right)}\, .
\label{eq:muComplexity}
\ee
For fixed temperature, the additional term $\ln|\omega/\mu|$ modifies the behavior of the complexity at both high $\omega \gg \mu$ and low frequencies $\omega \ll \mu$, reflecting the additional complexity needed to establish spatial correlations in the target state (\ref{eq:PsiPurif}).




\subsection{Purification with Additional Ancillary Squeezing}
\label{sec:sub:AdditionalSqueeze}

A number of different purifications can be used to construct the complexity of purification (\ref{eq:CoP}) since the minimization procedure scans over the ancillary parameters, while preserving the reduced density matrix.
Ideally, we wish to find the simplest purification that leads the smallest complexity of purification of the thermal mixed state.
We have already shown that extending the purification of the thermal density matrix beyond that of a thermofield double (\ref{eq:TFD}) to include a squeezing angle (\ref{eq:TFDphi}) qualitatively changes the value and functional dependence of the complexity of purification on the temperature.
However, it is possible to generalize the purification further, and it is worth investigating whether including additional ancillary parameters further changes the result in a qualitative way.

The purification of the thermal density matrix as a pure state consisting of a two-mode squeezed state (\ref{eq:TFDsqueeze}) can be generalized by the action of a two-mode {\bf rotation operator}
\be
|\Psi\rangle_{\phi,\theta} = \hat {\mathcal R}(\theta)\, \hat S_{\rm sq}(r,\phi) | 0\rangle \otimes |0 \rangle_{\rm anc}\, ,
\label{eq:psiRotation}
\ee
where
\be
\hat {\mathcal R}(\theta) = \exp \left[-i\theta \left(\hat a^\dagger \hat a + \hat a^\dagger_{\rm anc} \hat a_{\rm anc}\right)\right]\, .
\ee
The rotation operator 
acts on the two-mode squeezed state as a shift $\phi \rightarrow \phi + \theta$ and thus is a redundant purification, so we can set $\theta = 0$.

Because the ancillary degrees of freedom are traced out in the process of obtaining the density matrix, we can perform additional transformations operating purely in ${\cal H}_{\rm anc}$.
In particular, we consider an \emph{additional} single-mode squeezing of the ancillary oscillator by the squeezing parameter and angle $r_{\rm anc}, \phi_{\rm anc}$
\be
|\Psi\rangle_{\phi,r_{\rm anc},\phi_{\rm anc}} = \hat S_{\rm anc}(r_{\rm anc},\phi_{\rm anc})\ \hat S_{\rm sq}(r,\phi) |0\rangle \otimes |0\rangle_{\rm anc}\, ,
\label{eq:stateExtraSqueeze}
\ee
where
\be
\hat S_{\rm anc}(r_{\rm anc},\phi_{\rm anc}) = \exp\left[-\frac{r_{\rm anc}}{2}\left(e^{-2i\phi_{\rm anc}}\hat a_{\rm anc}^2 - e^{2i\phi_{\rm anc}}\hat a_{\rm anc}^{\dagger 2}\right) \right]\, .
\label{eq:ancSqueeze}
\ee
We have now introduced two additional free parameters $r_{\rm anc},\phi_{\rm anc}$, which will need to include together with the two-mode squeezing angle $\phi$ in the minimization of the complexity (\ref{eq:CoP}).
This generalization is similar to the purification considered in \cite{MyersMixed}, which also considered an additional squeezing of the ancillary degrees of freedom (again, with vanishing squeezing angle).
However, because our two-mode purification mixes the physical and ancillary degrees of freedom through the two-mode squeezing angle, in addition to having an additional parameter $\phi_{\rm anc}$, the calculation of the wavefunction and the minimization of the complexity corresponding to (\ref{eq:stateExtraSqueeze}) is somewhat more involved.

We can construct the position-space wavefunction corresponding to (\ref{eq:stateExtraSqueeze}) as
\be
\Psi(q,q_{\rm anc})_{\phi,r_{\rm anc},\phi_{\rm anc}} = \langle q, q_{\rm anc}|\Psi\rangle = {\cal S}_{\rm anc}(r_{\rm anc},\phi_{\rm anc}) \Psi_{\rm sq}(q,q_{\rm anc})\, ,
\label{eq:psiExtraSqueeze}
\ee
where the single-mode ancillary squeeze operator in position-space takes the form
\be
{\cal S}_{\rm anc} = \exp \left\{ \frac{i r_{\rm anc}}{2}\left[(q_{\rm anc} p_{\rm anc} + p_{\rm anc} q_{\rm anc}) \cos(2\phi_{\rm anc}) + (\omega^{-1} p_{\rm anc}^2 - \omega q_{\rm anc}^2) \sin(2\phi_{\rm anc})\right] \right\}\, , \,
\label{eq:ancSqueezePos}
\ee
with $p_{\rm anc} = -i \partial_{q_{\rm anc}}$.
Using the auxiliary squeezing variables
\be
v \equiv r_{\rm anc} \cos 2\phi_{\rm anc} \,, \qquad u \equiv -\frac{ir_{\rm anc}}{2\omega} \sin 2\phi_{\rm anc} \,,
\ee
and rewriting (\ref{eq:ancSqueezePos}) using a BCH formula \cite{Van_Brunt_2015} as
\be
\mathcal{S}_{\rm anc} = e^{v q_{\rm anc} \partial_{q_{\rm anc}}} \, e^{\frac{e^{v}-1}{v} u \partial_{q_{\rm anc}}^2} \, e^{\frac{1-e^{-2v}}{2v}u \omega^2 q_{\rm anc}^2 } \,,
\ee
the position-space wavefunction (\ref{eq:psiExtraSqueeze}) takes the form
\be
\Psi(q,q_{\rm anc})_{\phi,r_{\rm anc},\phi_{\rm anc}} = \tilde{\mathcal{N}}  \exp\left\{-\frac{\omega}{2}\, \tilde A q^2 - \frac{\omega}{2}\, \t A_{\rm anc} q_{\rm anc}^2 - \omega \t B q q_{\rm anc} \right\} \, .
\label{eq:psiExtraSqueezeGaussian}
\ee
The Gaussian coefficients are
\begin{subequations}\label{eq:coff_A1A2A12}
	\begin{eqnarray}
		\t A &\equiv & A - \frac{B^2}{\left[A - (1-e^{-2v}) \frac{u}{v}\right] + \left[2(e^{v}-1) \frac{u}{v}\right]^{-1} }\,, \\
		\t A_{\rm anc} &\equiv & \frac{e^{2v} }{\left[A -  (1-e^{-2v}) \frac{u}{v}\right]^{-1} + \left[2(e^{v}-1) \frac{u}{v}\right] }\,, \\
		\t B &\equiv & \frac{B e^{v} }{1 + \left[A - (1-e^{-2v}) \frac{u}{v}\right] \left[2(e^{v}-1) \frac{u}{v}\right] }\,.
	\end{eqnarray}
\end{subequations}
where $A,B$ are as defined in (\ref{eq:gaussianAB}).
Notice that the wavefunction (\ref{eq:psiExtraSqueezeGaussian}) has a similar Gaussian form as (\ref{eq:psiSqueezed}), except now the coefficients of the physical position $q^2$ and ancillary position $q_{\rm anc}^2$ terms are no longer equal, as expected since we have performed an additional squeezing of the ancillary oscillator.
Notice that in the limit of vanishing ancillary squeezing $r_{\rm anc}\rightarrow 0$, the parameters reduce to their two-mode squeezing values $\t A, \t A_{\rm anc} \rightarrow A, \t B \rightarrow B$, while in the low-temperature limit $\beta \omega \gg 1$, $A\rightarrow 1, B \rightarrow 0$ as before, so that the physical and ancillary degrees of freedom decouple.

The wavefunction (\ref{eq:psiExtraSqueezeGaussian}) is again Gaussian in form, and can be written in terms of the normal mode frequencies
\begin{equation}\label{eq:OmegaPM}
	\Omega_{\pm} = \frac{1}{2}\left( \t A + \t A_{\rm anc}  \pm \sqrt{(\t A-\t A_{\rm anc})^2 + 4\t B^2} \right)\,.
\end{equation}
The corresponding complexity, including the ancillary squeezing, is then \cite{Jefferson,cosmology1}
\begin{equation}\label{eq:CExtraSqueeze}
\mathcal{C}_{\phi, r_{\rm anc},\phi_{\rm anc}} = \frac{1}{2} \sqrt{ \ln^2 \left| \Omega_+\right| + \ln^2\left| \Omega_-\right| + \arctan^2\frac{\text{Im } \Omega_+}{\text{Re } \Omega_+} + \arctan^2\frac{\text{Im } \Omega_-}{\text{Re } \Omega_-} } \,.
\end{equation}

\begin{figure}[t]
	\centering
	\includegraphics[width=.8\textwidth]{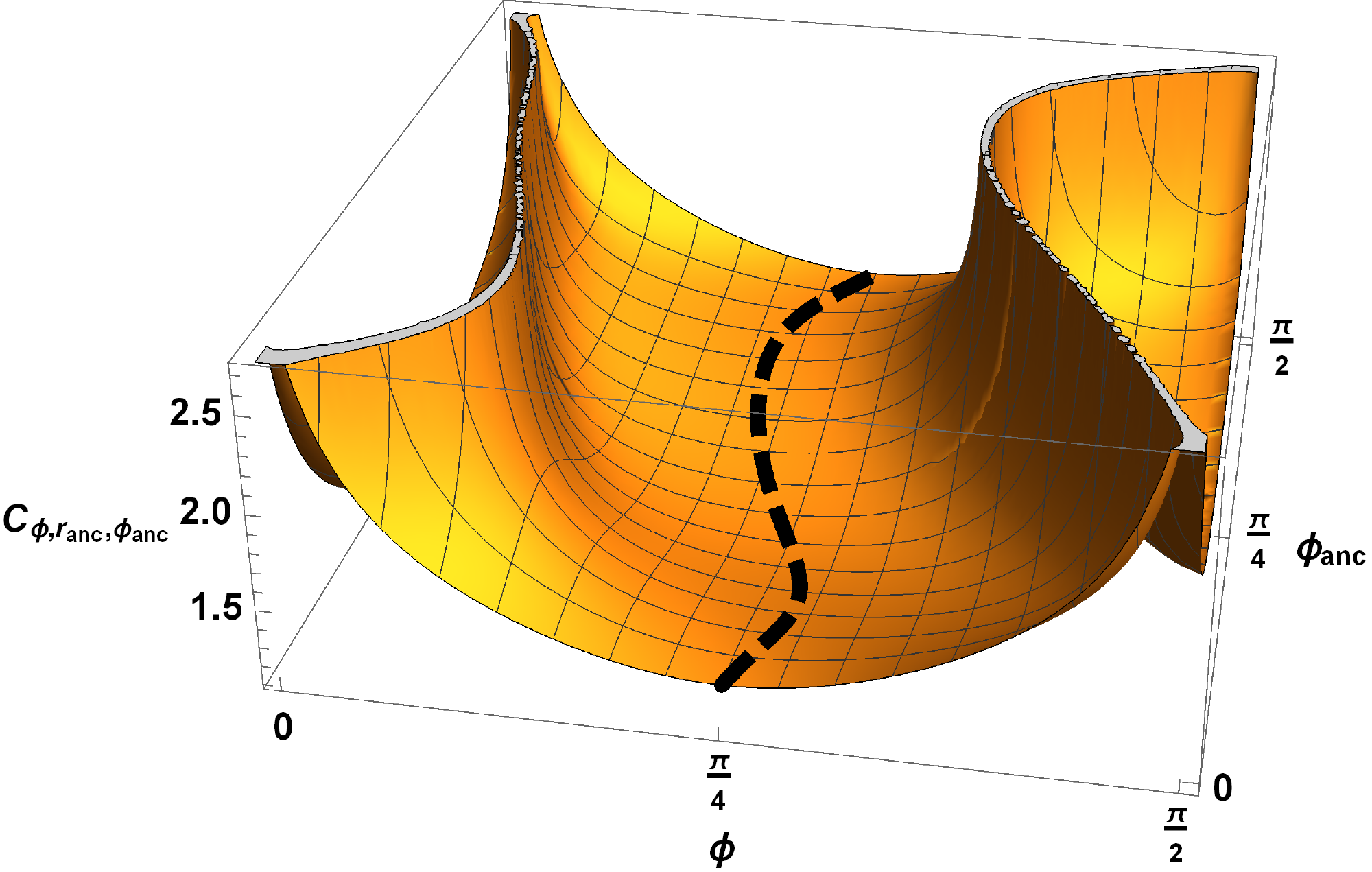}
	\caption{The purification complexity (\ref{eq:CExtraSqueeze}) as a function of the squeezing angles $(\phi,\phi_{\rm anc})$ for $\beta \omega = -2 \ln \tanh 50$ and ancillary squeezing $r_{\rm anc} = 1$. The thick dashed line corresponds to the minimum as a function of $(\phi,\phi_{\rm anc})$, tracing out a one-parameter curve in the valley of the complexity. In general, the minimized complexity of (\ref{eq:CExtraSqueeze}) over the free purification parameters $(\phi,r_{\rm anc},\phi_{\rm anc})$ is also a one-parameter curve, and approaches ${\cal C}_{\rm th} \approx \frac{\pi}{2\sqrt{2}}$ in the high-temperature limit.}
	\label{fig:ExtraMin}
\end{figure}

The thermal complexity of purification (\ref{eq:CoP}) is now the minimization of (\ref{eq:CExtraSqueeze}) over all three parameters $\phi, r_{\rm anc}, \phi_{\rm anc}$. Unfortunately, the expression is sufficiently complex that it is difficult to do a purely analytic minimization. 
Instead, we note that in the high-temperature limit $\beta \omega \ll 1$ the Gaussian parameters $\t A, \t A_{\rm anc}, \t B$ (\ref{eq:coff_A1A2A12}) and normal mode frequencies $\Omega_{\pm}$ (\ref{eq:OmegaPM}) become purely imaginary.
Further, we note that, as with the minimization of the complexity arising from two-mode squeezing (\ref{eqn:squeeze_vac_complexity2}), the minimum
of (\ref{eq:CExtraSqueeze}) occurs when the $\ln|\Omega_{\pm}|$ terms vanish, which together requires $\Omega_{\pm} = \pm i$.
This becomes a condition on the Gaussian parameters
\begin{equation}\label{eq:min_condition}
	\t A + \t A_{\rm anc} = 0 \,, \qquad  \frac{1}{4}(\t A-\t A_{\rm anc})^2 + \t B^2  = -1 \,.
\end{equation}
For example, the first condition implies that the minima are found as solutions to the relation
\begin{equation}
	\tan 2\phi_{\rm anc} + \frac{r_{\rm anc}}{2} \left(1- e^{-2v}\right) \tan 2\phi_{\rm anc} + \frac{\cot 2\phi_{\rm anc}}{ r_{\rm anc} (e^{v} -1)} = 0 \,.
	\label{eq:extraMinima}
\end{equation}
Together, the conditions (\ref{eq:min_condition}) impose two constraints on our three free parameters $\phi, r_{\rm anc},$ and $\phi_{\rm anc}$, implying that generically there is a one-parameter set of minima.
An example is shown in Figure \ref{fig:ExtraMin}, where a valley of minimum complexity for $r_{\rm anc} = 1$ is found as a curve in the $(\phi,\phi_{\rm anc})$ parameter space.
The corresponding complexity, in the high-temperature limit, is the same as for the two-mode squeezing
\be
{\cal C}_{\rm th} = \min_{\phi,r_{\rm anc},\phi_{\rm anc}} \mathcal{C}_{\phi, r_{\rm anc},\phi_{\rm anc}} = \frac{\pi}{2\sqrt{2}}\,.
\label{eq:extraCp}
\ee
As discussed above, in the low-temperature limit $\beta \omega \gg 1$ the physical and ancillary degrees of freedom decouple, so the extra squeezing of the ancillary degrees of freedom becomes irrelevant.

Altogether, we have found that generalizing the purification of the thermal density matrix to include an additional squeezing of the ancillary degrees of freedom as in (\ref{eq:stateExtraSqueeze}) leads to the \emph{same} minimized complexity of purification.
In particular, the minimized complexity of purification \emph{saturates} to a constant value (\ref{eq:extraCp}) at high temperature, and decays to zero at low temperature.
The additional degrees of freedom added lead to a one-parameter set of such minima, as opposed to the single $\phi = \pi/4$ minimized complexity of purification as with the pure two-mode squeezing.
Because the additional purification parameters $r_{\rm anc}, \phi_{\rm anc}$ do not change the value of the minimized complexity of purification, it seems that the additional freedom (and difficult) arising from their addition is undesirable.
However, there may be situations where boundary conditions do not permit a squeezing angle $\phi = \pi/4$, so that the additional freedom
of (\ref{eq:extraMinima}) is useful.
Nevertheless, the conceptual picture that arises from both the two-mode purification (\ref{eq:psiSqueezed}) and the additional ancillary squeezing (\ref{eq:stateExtraSqueeze}) for the thermal complexity of purification is the same: at low temperatures, the complexity vanishes exponentially as a function of the inverse temperature, while it saturates at high temperature to a constant value, becoming independent of the temperature.

\subsection{Complexity from Operator-State Mapping}
\label{sec:sub:OperatorState}


An alternative approach to assigning a pure state to the thermal density matrix $\hat \rho_{\rm th}$ (\ref{eq:thermalRho}) is the technique of \emph{operator-state mapping} (also known as \emph{channel-state mapping}) \cite{CHOI1975285,JAMIOLKOWSKI1972275}.
Consider an operator on our physical oscillator Hilbert space ${\cal H}$ with representation $\hat {\mathcal O} = \sum_{m,n} {\mathcal O}_{mn} |n\rangle\langle m|$. In its simplest form, the mapping associates a state $|{\mathcal O}\rangle$ to $\hat {\mathcal O}$ by flipping the bra to a ket,
\be
\hat {\mathcal O} = \sum_{m,n} {\mathcal O}_{mn} |n\rangle\langle m| \hspace{.2in} \longleftrightarrow \hspace{.2in}|{\mathcal O}\rangle = \frac{1}{\sqrt{\text{Tr} [\mathcal{O}^\dagger \mathcal{O}]}} \sum_{m,n} \mathcal{O}_{mn} | m \rangle \otimes | n \rangle_{\rm anc}\, .
\label{eq:OSOperatorMapping}
\ee
The state $|{\mathcal O}\rangle$ exists on the doubled Hilbert space ${\mathcal H}\otimes {\mathcal H}_{\rm anc}$, where again we denoted the extra copy of ${\mathcal H}$ as ${\mathcal H}_{\rm anc}$ to distinguish it from the original.
The process of operator-state mapping is superficially similar to purification studied in the previous subsection, in that both associate to the operator a state on a doubled Hilbert space.
One of the most important differences, however, is that the state $|\mathcal{O}\rangle$ in (\ref{eq:OSOperatorMapping}) associated to the operator $\hat {\mathcal O}$ is
\emph{unique} -- there are no free parameters introduced in the mapping (indeed, this is essential for the mapping to be an isomorphism) -- as compared to the purification (\ref{eq:TFDphi}), which introduces a squeezing angle $\phi$ that plays an important role in the minimization of the complexity.
The complexity associated with the operator-state mapping, in contrast, does not require a minimization over parameters.

For the thermal density matrix (\ref{eq:thermalRho}) we can then directly associate to it the two-mode squeezed state
\be
\hat \rho_{\rm th} = \left(1-e^{-\beta \omega}\right) \sum_{n} e^{-n\beta\omega} |n\rangle\langle n| \hspace{.2in} \longleftrightarrow \hspace{.2in} |\rho_{\rm th}\rangle = \frac{1}{\cosh^2 r} \sum_{n=0}^\infty (\tanh r)^{2n}\ |n\rangle \otimes |n\rangle_{\rm anc},\hspace{.2in} 
\label{eq:rhoOperatorState}
\ee
where $\tanh r = e^{-\beta \omega}$ is the two-mode squeezing parameter written in terms of the temperature as before.
It is now straightforward to write $|\hat \rho_{\rm th}\rangle$ in position space
\begin{equation}\label{eq:rho_q1q2}
	\begin{split}
		\rho(q,q_{\rm anc}) = \langle q,q_{\rm anc}|\rho_{\rm th}\rangle
		=
		\frac{(\omega/\pi)^{1/2}}{\sqrt{\cosh 2r}} \exp\left( -\frac{\omega}{2} {\mathcal A}(r) \left( q^2 + {q}_{\rm anc}^2 \right) + \omega {\mathcal B}(r) \, q\, q_{\rm anc}  \right) \,,
	\end{split}
\end{equation}
where
\begin{equation}
	{\mathcal A}(r) = \frac{1+\tanh^4 r}{1-\tanh^4 r} \, ,\qquad {\mathcal B}(r) = \frac{2 \tanh^2 r}{1-\tanh^4 r} \,.
\end{equation}
The corresponding {\bf operator-state thermal complexity} of (\ref{eq:rhoOperatorState}) relative to the ground state now follows directly from Appendix \ref{app:Complexity} with a vanishing squeezing angle $\phi = 0$
\be
{\mathcal C}_{\rm os}\left[\hat \rho_{\rm th}\right] = \frac{1}{\sqrt{2}} \ln \left(\frac{1+\tanh^2 r}{1-\tanh^2r}\right) = \frac{1}{\sqrt{2}} \ln \left(\frac{1+e^{-\beta\omega}}{1-e^{-\beta\omega}}\right)\, .
\label{eq:rhoOScomplexity}
\ee
For low-temperature $\beta \omega \gg 1$,(\ref{eq:rhoOScomplexity}) vanishes as ${\mathcal C}_{\rm os}[\hat \rho_{\rm th}]\sim e^{-\beta \omega}$, while for high temperatures $\beta \omega \rightarrow 0$ the complexity grows as ${\mathcal C}_{\rm os}[\hat \rho_{\rm th}] \sim 2 \ln (T/\omega)$.
The operator-state thermal complexity (\ref{eq:rhoOScomplexity}) thus does not saturate at high temperatures, as compared to the thermal complexity of purification (\ref{eq:Cp}).
The reason is clear: the operator-state mapping (\ref{eq:rhoOperatorState}) does not allow for the inclusion of a squeezing angle, so the state $|\rho_{\rm th}\rangle$ inherits a squeezing angle $\phi = 0$ from the thermal density matrix.
The operator-state thermal complexity is not identical to the $\phi = 0$ complexity of purification (\ref{eqn:squeeze_vac_complexity2}), however, since a factor of $\tanh^2r$ appears in (\ref{eq:rhoOperatorState}) instead of a factor of $\tanh r$, as in (\ref{eq:TFDsqueeze}).
This leads to a faster decay of the complexity ${\mathcal C}_{\rm os}\sim e^{-\beta \omega}$ at low-temperature as compared to the thermal complexity of purification.

An alternative method of computing the complexity of a mixed state density matrix such as $\rho_{\rm th}$ is
to work with the square root of the thermal density matrix $\hat \rho_{\rm th}$ as in \cite{haqueReducedDensityMatrix}, assigning to it the state
\be
|\rho_{\rm th}^{1/2}\rangle = \sum_{n} \left(\rho_{\rm th}^{1/2}\right)_{n n} |n\rangle \otimes |n\rangle_{\rm anc}\, ,
\label{eq:sqrtRhoOperatorState}
\ee
where $\left(\rho_{\rm th}^{1/2}\right)_{n n} = \langle n|\rho_{\rm th}^{1/2}|n\rangle$ are the matrix elements of $\hat \rho_{\rm th}^{1/2}$.
The original operator is then obtained as a trace over the ancillary degrees of freedom; for example,
\be
\hat \rho_{\rm th} &=& {\rm Tr}_{\rm anc}\left[|\rho_{\rm th}^{1/2}\rangle\langle \rho_{\rm th}^{1/2}|\right] 
	= {\rm Tr}_{\rm anc}\left[ \sum_{m,n} \left(\rho_{\rm th}^{1/2}\right)_{n n}\left(\rho_{\rm th}^{1/2}\right)_{m m} |n\rangle\langle m| \otimes |n\rangle_{\rm anc} \langle m |_{\rm anc}\right]  \\
	&=& \sum_n \left(\rho_{\rm th}\right)_{n n} |n\rangle \langle n|\, .
\ee
In order to calculate the complexity of the state (\ref{eq:sqrtRhoOperatorState}), we again write it as a Gaussian wavefunction in position-basis
\be
\langle q,q_{\rm anc}|\rho_{\rm th}^{1/2}\rangle = \rho_{\rm th}^{1/2}(q,q_{\rm anc}) = {\tilde {\mathcal N}} \exp\left( -\frac{\omega \alpha}{2} \left( q^2 + {q}_{\rm anc}^2 \right) + \omega \gamma \, q\, q_{\rm anc}  \right) \,.
\ee
where now
\be
	\alpha = {\mathcal A}(r) + {\mathcal B}(r) \,, \qquad \gamma = \sqrt{2{\mathcal B}(r) ({\mathcal A}(r) + {\mathcal B}(r))} \,,
\ee
and ${\mathcal A}(r), {\mathcal B}(r)$ are the same functions of the squeezing parameter given in (\ref{eq:rho_q1q2}).
It is straightforward to see that $\alpha, \gamma$ reduce to the purification quantities $A(r), B(r)$ with $\phi = 0$ from (\ref{eq:gaussianAB}), so
that the operator-state 
complexity corresponding to the state $\hat \rho_{\rm th}^{1/2}$ relative to the ground state reduces to
\begin{equation}\label{eqn:C_stateOP_map}
	\mathcal{C}_{\rm os}\left[\hat \rho_{\rm th}^{1/2}\right] = \sqrt{2} \, \text{arctanh}\left( e^{-\beta\omega/2} \right) \,.
\end{equation}
Thus, while the low-temperature behavior of $\mathcal{C}_{\rm os}\left[\hat \rho_{\rm th}^{1/2}\right] \sim e^{-\beta \omega/2}$ matches that of the minimal complexity of purification, the high-temperature behavior $\mathcal{C}_{\rm os}\left[\hat \rho_{\rm th}^{1/2}\right] \sim \ln(T/\omega)$ grows with temperature, rather than saturating.

\begin{figure}[t]
	\centering
	\includegraphics[width=.6\textwidth]{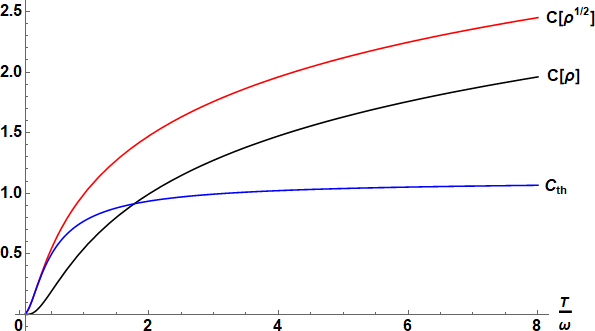}
	\caption{A complexity can be associated with a mixed thermal density matrix $\hat \rho_{\rm th}$ through the complexity of purification ${\mathcal C}_{\rm th}$ (\ref{eq:CoP}), operator-state complexity of the density matrix ${\mathcal C}_{\rm os}[\hat \rho_{\rm th}]$ (\ref{eq:rhoOScomplexity}), and the operator-state complexity of the square root of the density density matrix ${\mathcal C}_{\rm os}[\hat \rho_{\rm th}^{1/2}]$ (\ref{eqn:C_stateOP_map}). As also discussed in the previous section, the complexity of purification saturates at high temperatures, while the operator-state complexities grow logarithmically with temperature $\sim \ln (T/\omega)$.}
	\label{fig:C_stateOP}
\end{figure}

Figure \ref{fig:C_stateOP} compares the results for the complexity of the thermal density matrix $\hat \rho_{\rm th}$ for the different techniques we have used in this section: the minimal complexity of purification ${\mathcal C}_{\rm th}$ (\ref{eq:Cp}), the operator-state complexity associated with the density matrix ${\mathcal C}_{\rm os}[\hat \rho_{\rm th}]$ (\ref{eq:rhoOScomplexity}), and the operator-state complexity associated with the square root of the thermal density matrix ${\mathcal C}_{\rm os}[\hat \rho_{\rm th}^{1/2}]$ (\ref{eqn:C_stateOP_map}).
Each technique gives a complexity that vanishes exponentially with the temperature at low temperatures, while at high temperatures their behaviours differ: the minimal complexity of purification saturates at high temperatures, while the operator-state complexities grow as $\ln (T/\omega)$.
Because of the latter behavior, the operator-state complexity diverges in the infrared logarithmically with the IR cutoff.

Qualitatively, each of the states $|\Psi\rangle_\phi$ (\ref{eq:TFDphi}), $|\rho_{\rm th}\rangle$ (\ref{eq:rhoOperatorState}) and $|\rho_{\rm th}^{1/2}\rangle$ reproduce the thermal expectation values in the physical Hilbert space when traced over the additional degrees of freedom, so which complexity should we associate with the thermal density matrix $\hat \rho_{\rm th}$?
Viewing complexity as the minimization of the number of gates (``circuit depth'') needed to construct the target state from a given reference state, 
the {\bf complexity of purification} allows us to construct a state, namely $|\Psi\rangle_{\phi = \pi/4}$, that reproduces all of our physical observables with a smaller number of gate resources than the other techniques.
In contrast, because it maps to a unique target state, the operator-state complexity for either $\hat \rho_{\rm th}$ or $\hat \rho_{\rm th}^{1/2}$ picks out a unique path in state space, which has no particular reason to be minimal (and in fact we find it is not minimal).
Thus, from the perspective of resource management, the {\bf complexity of purification} allows us to simulate the mixed state as a pure state with a smaller number of resources.



\section{Thermal Complexity in Curved Spacetimes}
\label{sec:Curved}

In the previous section we have shown how the complexity for a thermal density matrix of a harmonic oscillator can be obtained by calculating the complexity of a corresponding two-mode squeezed state parameterized by a squeezing parameter (fixed by the temperature) and a squeezing angle.
Since the thermal density matrix is independent of the squeezing angle, we minimized the complexity with respect to the squeezing angle and found that the minimum complexity corresponds to a non-zero squeezing angle.

The thermal density matrix of a harmonic oscillator arises as a simple model in many different contexts; here, we would like to explore the complexity of thermal density matrices that arise naturally when considering quantum fields on curved spacetimes.


\subsection{Complexity and Hawking Radiation}
\label{sec:sub:Hawking}

Thermal density matrices naturally arise when considering quantum fields on spacetimes with horizons through the Unruh effect and the generation of Hawking radiation, so these are ideal backgrounds to consider applications of the thermal complexity of purification.
See \cite{BirrellDavies,Ford:1997hb,Jacobson:2003vx,UnruhEffectReview,krishnan2010quantum,Polchinski_2016} for some useful reviews.

\begin{figure}[t]\centering
	\begin{subfigure}[b]{0.56\textwidth}
		\centering\includegraphics[width=\textwidth]{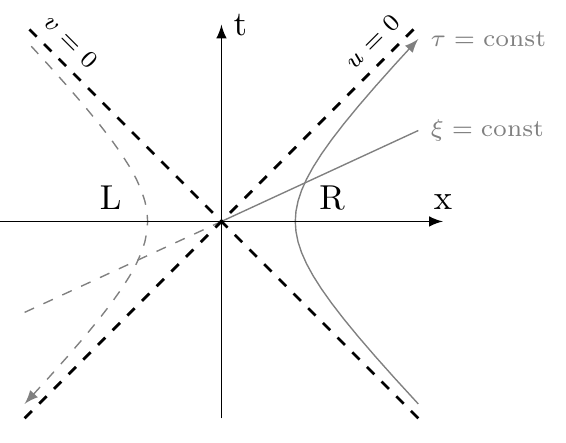}
		\caption{Rindler spacetime split into ${\rm R}$ and ${\rm L}$ wedges.}
		\label{subfig:Rindler}
	\end{subfigure}
	\hfill
	\begin{subfigure}[b]{0.43\textwidth}
		\centering\includegraphics[width=\textwidth]{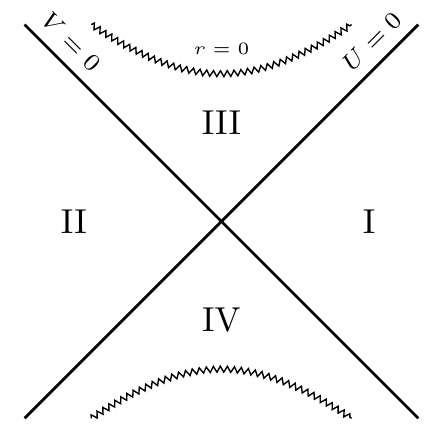}
		\caption{Kruskal spacetime for a black hole.}
		\label{subfig:Kruskal}
	\end{subfigure}
	\caption{Spacetime diagrams for Rindler and Kruskal spaces.}
	\label{fig:SpacetimeDiagrams}
\end{figure}

As an illustrative example that contains most of the relevant details, let us begin by considering $(1+1)$-dimensional Minkowski space as seen by a uniformly accelerating observer.
The metric can be written
\be
ds^2 = -dt^2 + dx^2 = -du\ dv=e^{2a\xi}(d\tau^2 - d\xi^2)\, ,
\ee
where $u = t-x, v = t+x$ are Minkowski light-cone coordinates and $(\tau,\xi)$ defined by
\be
t = \frac{1}{a} e^{a\xi} \sinh(a\tau)\, ,\hspace{.2in} x = \frac{1}{a} e^{a\xi} \cosh(a\tau)
\label{eq:RindlerCoords}
\ee
are coordinates adapted to a uniformly accelerating observer with acceleration $a$ along $\xi = 0$.
The coordinates $(\tau,\xi)$ in (\ref{eq:RindlerCoords}) only cover part of the original Minkowski space, the so-called right $\rm R$ Rindler wedge $|x| > t$;
a similar definition is needed, with additional minus signs, for the $\rm L$ wedge. 
An interesting and important feature of Rindler coordinates is that $u = 0$ is a future horizon for an observer traveling along a constant $\xi$ line in the $\rm R$ wedge ($v=0$ is correspondingly a past horizon), so that the $\rm R$ and $\rm L$ patches are causally disconnected from each other. 
See Figure \ref{subfig:Rindler} for a spacetime diagram of Minkowski space and ${\rm R,L}$ Rindler wedges.

A massless scalar field $\phi$ on this spacetime can be expanded in Minkowski $(t,x)$ plane wave modes
\be
\hat \phi = \sum_k \left(u_k^M\ \hat a_k + u_k^{M*}\ \hat a_k^\dagger\right)
\ee
where the Minkowski vacuum is defined by the annihilation operator $\hat a_k |0\rangle_M = 0$.
We can also expand the scalar field in modes adapted to Rindler $(\tau,\xi)$ coordinates
\be
\hat \phi = \sum_k \left(v_k^{\rm R}\ \hat b_k^{\rm R} + v_k^{\rm L}\ \hat b_k^{\rm L} + v_k^{\rm R*}\ \hat b_k^{\rm R\dagger} + v_k^{\rm L*}\ \hat b_k^{\rm L\dagger}\right)
\label{eq:RindlerModeExpansion}
\ee
where $\hat v^{\rm R,L}_k$ are only non-zero in the ${\rm R,L}$ wedges, respectively, and
the Rindler vacuum can be written as the direct product of vacuum states on the left and right wedges $|0\rangle_{\rm R} \otimes |0\rangle_{\rm L}$, which are annihilated by the left- and right-Rindler annihilation operators, $\hat b_k^{\rm R} |0\rangle_{\rm R} =0= \hat b_k^{\rm L} |0\rangle_{\rm L}$.

The Minkowski and Rindler vacuum states are not equivalent;
instead, the Minkowski vacuum of a single mode $k$ of the scalar field $\phi$ as seen in the basis of Rindler modes (\ref{eq:RindlerModeExpansion}) takes the form of an two-mode squeezed state entangling the ${\rm L}$ and ${\rm R}$ modes \cite{UnruhEffectReview,krishnan2010quantum}, which we will call the ``Unruh'' state
\be
|0\rangle_M = \left(1-e^{-2\pi k/a}\right)^{1/2} \sum_{n_k=0}^\infty e^{-\frac{n\pi k}{a}}\ |n_k\rangle_{\rm R} \otimes |n_k\rangle_{\rm L}\, .
\label{eq:MinkSqueezed}
\ee
Comparing (\ref{eq:MinkSqueezed}) with (\ref{eq:TFDsqueeze}), the squeezing parameter is $\tanh r = e^{-\pi k/a}$ with squeezing angle $\phi = \pi/2$.
The corresponding complexity of this Unruh state relative to a Rindler ground state $|0\rangle_{\rm R}\otimes |0\rangle_{\rm L}$ is
\be
{\mathcal C}_{\rm Unruh}(k) = \frac{1}{\sqrt{2}} \ln \left(\frac{1+e^{-\frac{\pi k}{a}}}{1-e^{-\frac{\pi k}{a}}}\right)\, ,
\label{eq:MinkComplexity}
\ee
and grows logarithmically ${\mathcal C}_{\rm Unruh} \sim r\sim \ln(a/k)$ for large accelerations $a/k \gg 1$.
Interestingly, the complexity (\ref{eq:MinkComplexity}) of the Unruh two-mode squeezed state is \emph{maximal} with respect to the squeezing angle -- that is, for fixed acceleration, the squeezing angle
of the squeezed state (\ref{eq:MinkSqueezed}) chooses the largest possible complexity. We will return to this observation at the end of this section.

A Rindler observer in the $\rm R$-wedge sees a density matrix with the unobservable $\rm L$ modes traced out
\be
\hat \rho_{\rm R} = {\rm Tr}_{\rm L} \left[|0\rangle_M \langle 0|_M\right]
= \left(1-e^{-2\pi k/a}\right) \sum_{n_k} e^{-2\pi n_k k/a}\ |n_k\rangle_{\rm R}\ \langle n_k|_{\rm R}\, ,
\label{eq:RindlerThermalrho}
\ee
which is identified as a thermal density matrix of Hawking radiation with temperature $T = a/2\pi$.
The thermal complexity of purification from Section \ref{sec:sub:TwoModePurif} assigns a minimal purification to (\ref{eq:RindlerThermalrho})
\be
|\Psi\rangle_{\rm Hawk,p} = \left(1-e^{-2\pi k/a}\right)^{1/2} \sum_{n_k=0}^\infty (-1)^{n_k} e^{-2in_k\phi} e^{-\frac{\pi n_k k}{a}}\ |n_k\rangle_{\rm R}\ \otimes |n_k\rangle_{\rm L'}\, ,
\label{eq:RindlerPurification}
\ee
with the same squeezing parameter $r$ as with the Unruh state, but now with squeezing angle $\phi = \pi/4$ as determined by the minimization of the complexity of purification.
Note that the purification includes an ancillary Hilbert space, which we will denote as $\rm L'$ due to its similarity with the $\rm L$ wedge of Rindler space.
Comparing the Unruh squeezed state (\ref{eq:MinkSqueezed}) and the purification of the Hawking radiation (\ref{eq:RindlerPurification}), these two purifications
differ only by the value of the squeezing angle $\phi$:
The Unruh squeezed state (\ref{eq:MinkSqueezed}) has $\phi = \pi/2$, which as discussed in Section \ref{sec:sub:TwoModePurif} maximizes the corresponding complexity for fixed acceleration, while the minimization arising from the thermal complexity of purification selects $\phi = \pi/4$. The minimized complexity of purification for the Hawking thermal density matrix (\ref{eq:RindlerThermalrho}) is instead
\be
{\mathcal C}_{\rm Hawk,p}(k) = \frac{1}{\sqrt{2}} \left|\arctan\left(2 \frac{e^{-\pi k/a}}{1-e^{-2\pi k/a}}\right)\right|\, ,
\label{eq:RindlerComplexity}
\ee
which saturates to ${\mathcal C}_{\rm Hawk,p} \sim \pi/(2\sqrt{2})$ for large accelerations $a/k \gg 1$.
The minimization imposed by the thermal complexity of purification leads to a different qualitative behavior of the complexity of the Hawking radiation as a function of the acceleration, compared to the complexity of the full Unruh squeezed state.
Nevertheless, because the purification process preserves expectation values, both purifications (\ref{eq:MinkSqueezed}), (\ref{eq:RindlerPurification}) of the thermal density matrix (\ref{eq:RindlerThermalrho}) are equivalent to an observer in the $R$ wedge.
Thus, a $\rm R$ wedge observer can reconstruct thermal expectation values by using the purification (\ref{eq:RindlerPurification}), which is 
easier to build (in that it has a smaller complexity) compared to the Unruh squeezed state (\ref{eq:MinkSqueezed}).

The role of the squeezing angle arises in the analytic continuation of positive frequency mode functions from the $\rm R$ wedge to the $\rm L$ wedge.
To see this, we
expand the global scalar field in a set of different positive frequency mode functions in Minkowski space 
\be
\hat \phi = \sum_k\left(\bar U_k^I \hat c_k^I + \bar U_k^{II} \hat c_k^{II} + \bar U_k^{I*} \hat c_k^{I\dagger} + \bar U_k^{II*} \hat c_k^{II\dagger}\right)\, 
\ee
where the modes $\bar U_k^I, \bar U_k^{II}$ are defined as
combinations of the Rindler mode functions that have positive frequency in Minkowski coordinates.
This corresponds to a Bogoliubov transformation between the Minkowski modes $\hat c_k^I, \hat c_k^{II}$ and the Rindler modes $\hat b_k^{\rm R}, \hat b_k^{\rm L}$
\be
\hat c_k^I = \sqrt{\frac{e^{\frac{\pi k}{a}}}{2 \sinh(\frac{\pi k}{a})}}\left(\hat b_k^{\rm R} + e^{-\frac{\pi k}{a}} \hat b_k^{\rm L\dagger}\right) \, ;\hspace{.2in}
\hat c_k^{II} = \sqrt{\frac{e^{\frac{\pi k}{a}}}{2 \sinh(\frac{\pi k}{a})}}\left(\hat b_k^{\rm L} + e^{-\frac{\pi k}{a}} \hat b_k^{\rm R\dagger}\right)\, .\label{eq:RindlerBogo}
\ee
The factors of $e^{-\frac{\pi \omega}{a}}$ arise from the analytic continuation of the positive frequency mode functions from the $\rm R$ wedge to the $\rm L$ wedge, and give rise to the temperature as seen by an $\rm R$ wedge Rindler observer.

As reviewed in Appendix \ref{app:Bogo}, the Bogoliubov transformations (\ref{eq:RindlerBogo}) can be seen as a special case of a two-mode squeezing transformation
\be
\hat c_k^I = \sqrt{\frac{e^{\frac{\pi k}{a}}}{2 \sinh(\frac{\pi k}{a})}}\left(\hat b_k^{\rm R} - e^{2i\phi} e^{-\frac{\pi k}{a}} \hat b_k^{\rm L'\dagger}\right)\, ; \hspace{.2in}
\hat c_k^{II} = \sqrt{\frac{e^{\frac{\pi k}{a}}}{2 \sinh(\frac{\pi k}{a})}}\left(\hat b_k^{\rm L'} - e^{2i\phi} e^{-\frac{\pi k}{a}} \hat b_k^{\rm R\dagger}\right)\, ,\label{eq:SqueezeBogo}
\ee
with a squeezing angle $\phi = \pi/2$ that
ensures continuity of the $\bar U_k^I, \bar U_k^{II}$ mode functions in Minkowski space across the $u = 0 = v$ boundary between the $\rm R$ and $\rm L$ wedges.
Interpreting the $L'$ wedge in a similar geometric way, an arbitrary squeezing angle $\phi$ for the purification (\ref{eq:RindlerPurification}) would instead imply a phase discontinuity when crossing the Rindler horizon, which can be absorbed in the Rindler mode function behind the horizon, 
$v_k^{\rm L'} \rightarrow v_k^{\rm L'} e^{2i\phi}$.
The minimized squeezing angle $\phi = \pi/4$ arising from the complexity of purification thus results in a purely imaginary phase.
It is interesting that this phase in the Bogoliubov transformations (\ref{eq:SqueezeBogo}) has qualitatively different effects on the complexity of the resulting squeezed state, and it would be interesting to see whether this phase has any other physical effects.

Other spacetimes with horizons also exhibit the Unruh effect, for similar conceptual and technical reasons.
For example, consider a black hole spacetime, as shown in the Kruskal diagram of Figure \ref{subfig:Kruskal}.
Ignoring the angular directions and treating spacetime as $(1+1)$-dimensional, the metric is
\be
ds^2 &=& -\left(1-\frac{2GM}{r}\right) dt^2 + \left(1-\frac{2GM}{r}\right)^{-1} dr^2 = -\left(1-\frac{2GM}{r}\right) du\ dv \nonumber \\
&=& -\frac{16 G^2M^2}{r} e^{-r/2GM}\ dU\ dV
\ee
where we introduced the lightcone coordinates
\be
u = t - r_* = -4GM \ln(-U/2GM), \hspace{.2in} v = t+r_* = 4 GM\ln(V/2GM)
\ee
where $r_* = r+2GM\ln(r-2GM)$.
A massless scalar field can be expanded in modes
\be
\hat \phi &=& \sum_k \left(u_k\ \hat a_k + u_k^* \hat a_k^\dagger\right) \\
&=& \sum_k \left(v_k^{\rm I}\ \hat b_k^{\rm I} +v_k^{\rm II}\ \hat b_k^{\rm II}+ v_k^{\rm I*} \hat b_k^{\rm I\dagger} +v_k^{\rm II*} \hat b_k^{\rm II\dagger} \right)
\ee
where the $u_k$ modes are defined with respect to the global $(U,V)$ coordinates and define the vacuum in the asymptotic past $\hat a_k |0\rangle_{\rm past} = 0$, and the $v_k^{\rm I,II}$ modes are defined with respect to the $(u,v)$ coordinates in the ${\rm I, II}$ patches and define the vacuum seen by an observer outside the black hole in the asymptotic future $\hat b_k^{\rm I} |0\rangle_{\rm I} = 0$ and the internal vacuum with respect to the modes inside the black hole $\hat b_k^{\rm II}|0\rangle_{\rm II} = 0$.
As in Rindler space, these modes are mixed by a Bogoliubov transformation (again, see \cite{BirrellDavies,Ford:1997hb,Jacobson:2003vx,UnruhEffectReview,krishnan2010quantum,Polchinski_2016} for reviews), so that the past vacuum can be written as a squeezed state with $\phi = \pi/2$, entangling the external and internal modes with each other as an Unruh state
\be
|0\rangle_{\rm past} &=& {\mathcal N}_k\ {\rm exp} \left(e^{-4\pi GM k}\hat b_k^{\rm I\dagger} \hat b_k^{\rm II \dagger}\right) |0\rangle_{\rm I}\otimes |0\rangle_{\rm II} \nonumber \\
	&=& {\mathcal N}_k\ \sum_{n_k} e^{-4\pi GM k\ n_k }\ |n_k\rangle_{\rm I}\otimes |n_k\rangle_{\rm II}\, ,
	\label{eq:BHSqueezed}
\ee
with corresponding complexity
\be
{\mathcal C}_{\rm Unruh}(k) = \frac{1}{\sqrt{2}} \ln \left(\frac{1+e^{-4\pi GM k}}{1-e^{-4\pi GM k}}\right)\, .
\label{eq:BHComplexity}
\ee

As with Rindler space, the density matrix as seen by an external observer in the asymptotic future is obtained from (\ref{eq:BHSqueezed}) by
tracing out the modes inside the horizon, leading to a thermal density matrix with temperature $T_{BH} = (8\pi G M)^{-1}$ corresponding to Hawking radiation of the scalar field $\phi$.
The corresponding thermal purification of complexity associates a squeezed state to this density matrix, entangling the external modes associated with ${\rm I}$ with ancillary degrees of freedom on a space ${\rm II'}$ with squeezing angle $\phi = \pi/4$, leading to the purification complexity
\be
{\mathcal C}_{\rm Hawk,p}(k) = \frac{1}{\sqrt{2}} \left|\arctan\left(2 \frac{e^{-4\pi GM k}}{1-e^{-8\pi GM k}}\right)\right|\, ,
\label{eq:BHPurifComplexity}
\ee
As before, the complexity (\ref{eq:BHPurifComplexity}) associated with the purification of the thermal density matrix
is qualitatively different from that of the global Unruh state complexity (\ref{eq:BHComplexity}), particularly at low-frequencies or small masses, saturating
instead of growing as $GM k \rightarrow 0$.

We have reviewed how quantum field theory on curved spacetimes with a horizon naturally leads to a Unruh two-mode squeezed state entangling degrees of freedom on either side of the horizon through the Unruh effect, with an associated complexity.
Interestingly, the complexity of this Unruh squeezed state is {\bf maximal} with respect to the squeezing angle. In particular, the squeezing angle that naturally arises in the Bogoliubov transformation of the Unruh state gives the largest complexity for a given squeezing, such that the complexity grows with the squeezing ${\mathcal C}_{\rm Unruh} \sim r$, which itself is an increasing function of the acceleration (for Rindler spacetimes) or the black hole mass.
For a black hole, this agrees with expectations from other work that black holes are maximally chaotic quantum systems \cite{kitaev,FastScramblers,Maldacena2016-mb}, and perhaps similar statements apply to Rindler space as well.
It is therefore interesting to speculate that the complexity for the Unruh black hole state is maximal because of some deeper principle that applies more generally, which may take the form of a tendency for the field configuration to maximize the complexity in a kind of second law of complexity, similar to entropy. At least in our simple model of a scalar field on curved backgrounds, this seems to be the case.

However, an observer outside the horizon sees a thermal density matrix of Hawking radiation, obtained by tracing out over the internal modes.
The thermal complexity of purification from Section \ref{sec:sp_purifcn} then associates an ancillary two-mode squeezed state description to the Hawking radiation.
The purified Hawking radiation state takes a similar form as the Unruh squeezed state, but with a squeezing angle that instead {\bf minimizes} the complexity at a constant value ${\mathcal C}_{\rm Hawk,p}\sim \pi/(2\sqrt{2})$.
This implies that for an observer outside of the horizon, it is easier to build the Hawking radiation as a two-mode squeezed state entangling the Hawking radiation with ancillary degrees of freedom at a particular squeezing angle $\phi = \pi/4$ that differs from the Unruh state.
For a black hole of mass $M$, the difference between the maximal Unruh complexity and the minimal purification complexity for fixed frequency $k$ is only weakly dependent on the mass of the black hole, ${\mathcal C}_{\rm Unruh}/{\mathcal C}_{\rm p} \sim \ln (GMk)$.
However, including many frequencies $k$ this effect can potentially become an important effect.

Finally, as discussed in Section \ref{sec:sub:TwoModePurif}, note that the Unruh state (\ref{eq:MinkSqueezed}) and the purification of the Hawking radiation (\ref{eq:RindlerPurification}) have identical entanglement entropies that only depend on the temperature and frequency through $\beta k$
\be
S_{\rm en} = \ln \left(1-e^{-\beta k}\right) - \frac{e^{-\beta k}}{1-e^{-\beta k}} \ln\left(e^{-\beta k}\right)\, .
\ee
Thus, while there are several ways that an observer outside the horizon can construct a pure state representing the Hawking radiation with identical entanglement entropies (parameterized by different values of the squeezing angle $\phi$), the purification
(\ref{eq:RindlerPurification}) is minimal with respect to the complexity of building the state, by a factor $\sim \ln (GMk)$ compared
to the usual Unruh state (\ref{eq:MinkSqueezed}) for black holes.
It would be interesting to consider whether this difference has an impact on the information contained in the Hawking radiation in a more robust treatment.
We leave these interesting questions for future work.

\subsection{Complexity and Cosmological Perturbations}
\label{sec:sub:SqueezedCosmo}

Squeezed states and thermal density matrices also arise in the quantum description of cosmological perturbations, 
in which the time-dependence of the metric induces the pair creation of particles from the vacuum.

We will briefly review the description of cosmological perturbations as squeezed states; see \cite{Mukhanov,Grishchuk,Albrecht,Martin1,Martin2,cosmology1,cosmology2} for more details.
Our metric is the spatially flat Friedmann-Lemaitre-Robertson-Walker (FLRW) metric
\begin{equation}
	ds^2 = -dt^2 + a(t)^2 d\vec{x}^2 = a(\eta)^2 \left(-d\eta^2+d\vec{x}^2\right)\, .
\end{equation}
The Hubble expansion rate of this background is denoted by $H = \dot a/a$, where a dot denotes a derivative with respect to cosmic time $t$.
On this background we will consider fluctuations of a scalar field, which combine with fluctuations of the metric to
form the gauge-invariant curvature perturbation ${\mathcal R}$.
Written in terms of the
Mukhanov-Sasaki variable
$v \equiv z {\mathcal R}$
where $z \equiv a\, \sqrt{2\epsilon}$, with $\epsilon = -\dot H/H^2 = 1-{\mathcal H}'/{\mathcal H}^2$,
the action takes the simple form
\be
S = \frac{1}{2} \int d\eta\, d^3x \left[v'^2 - (\partial_i v)^2 + \left(\frac{z'}{z}\right)^2 v^2 - 2 \frac{z'}{z} v' v\right]\, .
\label{CosmoAction}
\ee
where a prime denotes a derivative with respect to conformal time and ${\mathcal H} = a'/a$.
This action represents perturbations of a free scalar field coupled to an external time-varying source. Promoting the perturbation to a quantum field and expanding into creation and annihilation modes
\be
\hat v(\eta,\vec{x}) = \int \frac{d^3k}{(2\pi)^{3/2}} \frac{1}{\sqrt{2k}} \left(\hat c_{\vec{k}}e^{i\vec{k}\cdot \vec{x}} + \hat c_{-\vec{k}}^\dagger e^{i\vec{k}\cdot \vec{x}} \right)\, \, ,
\ee
the Hamiltonian can be written as
\be
\hat H = \int d^3k\, \hat {\mathcal H}_{\vec{k}} = \int d^3k \left[k\left(\hat c_{\vec{k}} \hat c_{\vec{k}}^\dagger + \hat c_{-\vec{k}}^\dagger \hat c_{-\vec{k}}\right) 
- i \frac{z'}{z} \left(\hat c_{\vec{k}} \hat c_{-\vec{k}} - \hat c_{\vec{k}}^\dagger \hat c_{-\vec{k}}^\dagger\right)\right]\, .
\label{CosmoH}
\ee
The momentum structure of the Hamiltonian indicates that the interaction with the background leads to particle creation in pairs with opposite momenta. Because of this, we are naturally led to consider our states as appearing in two-mode pairs $(\vec{k},-\vec{k})$.

\begin{figure}[t]
	\begin{subfigure}[b]{0.49\textwidth}
		\includegraphics[width=\textwidth]{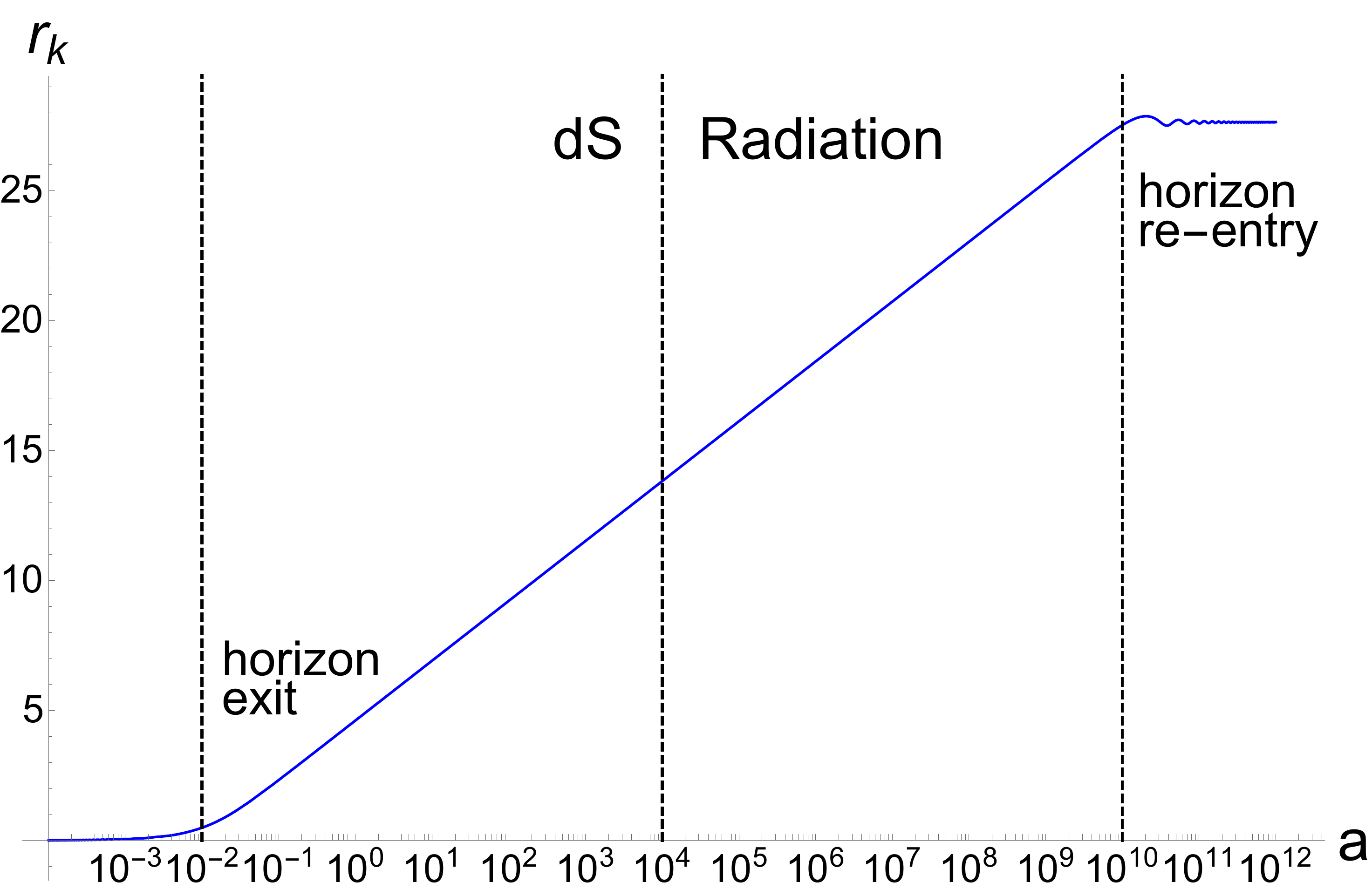}
		\caption{Squeezing parameter $r_k$.}
	\end{subfigure}\hspace{.1in}
	\begin{subfigure}[b]{0.49\textwidth}
		\includegraphics[width=\textwidth]{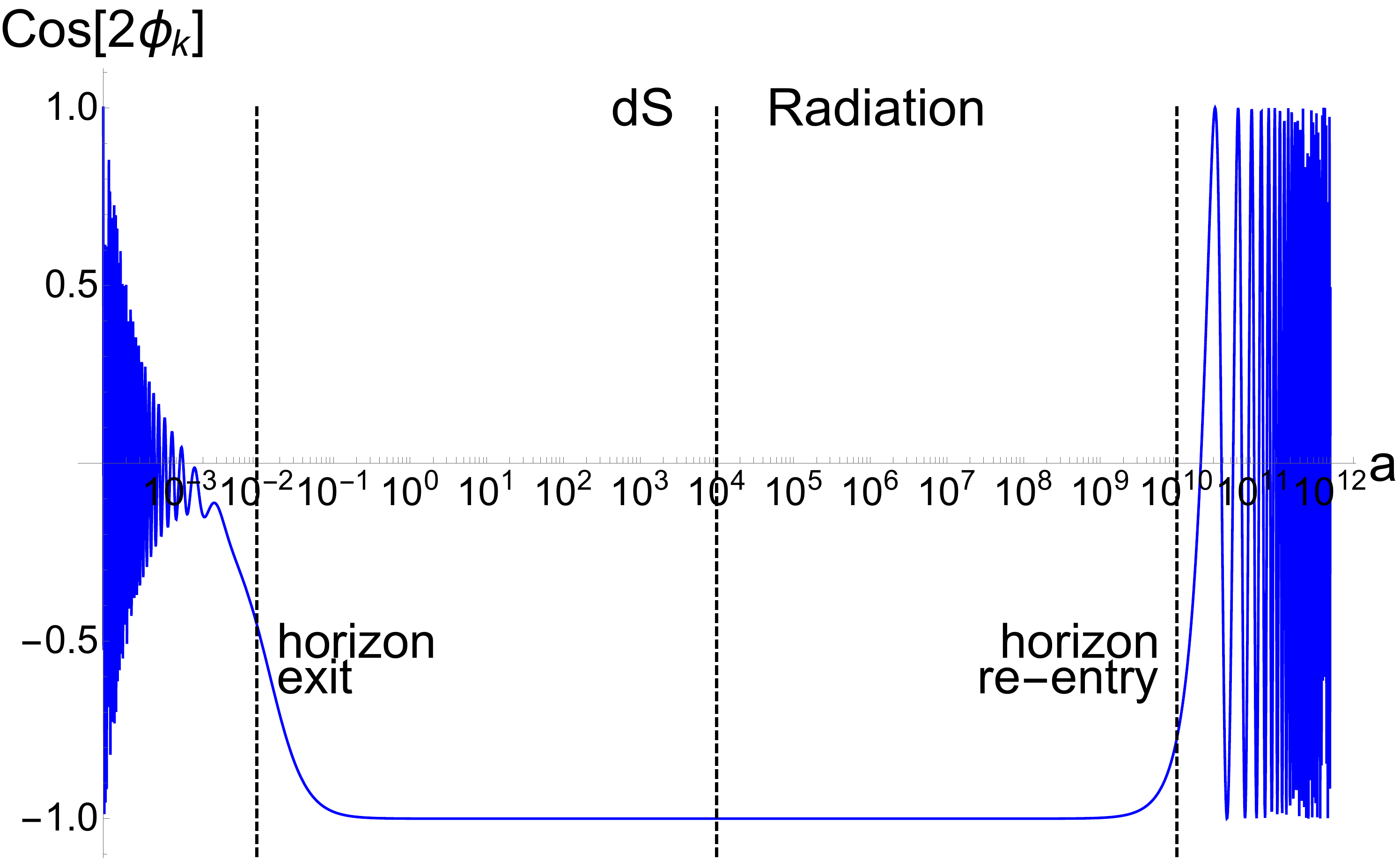}
		\caption{Squeezing angle $\phi_k$.}
	\end{subfigure}
\caption{The squeezing parameter and squeezing angle of a pure state cosmological perturbation (\ref{psi1}) with fixed $k$ are shown for a simple model of the early Universe consisting of a period of de Sitter accelerated expansion, followed by radiation domiation. The squeezing parameter $r_k$ grows while the mode is outside the horizon, while $\phi_k\approx -\pi/2$ there. When the mode eventually re-enters the horizon the squeezing parameter ``freezes in'', while the squeezing angle begins to grow, causing rapid oscillations in $\cos (2\phi_k)$.}
\label{fig:CosmoSqueezing}
\end{figure}

\begin{figure}[t]
	\centering \includegraphics[width=0.95\textwidth]{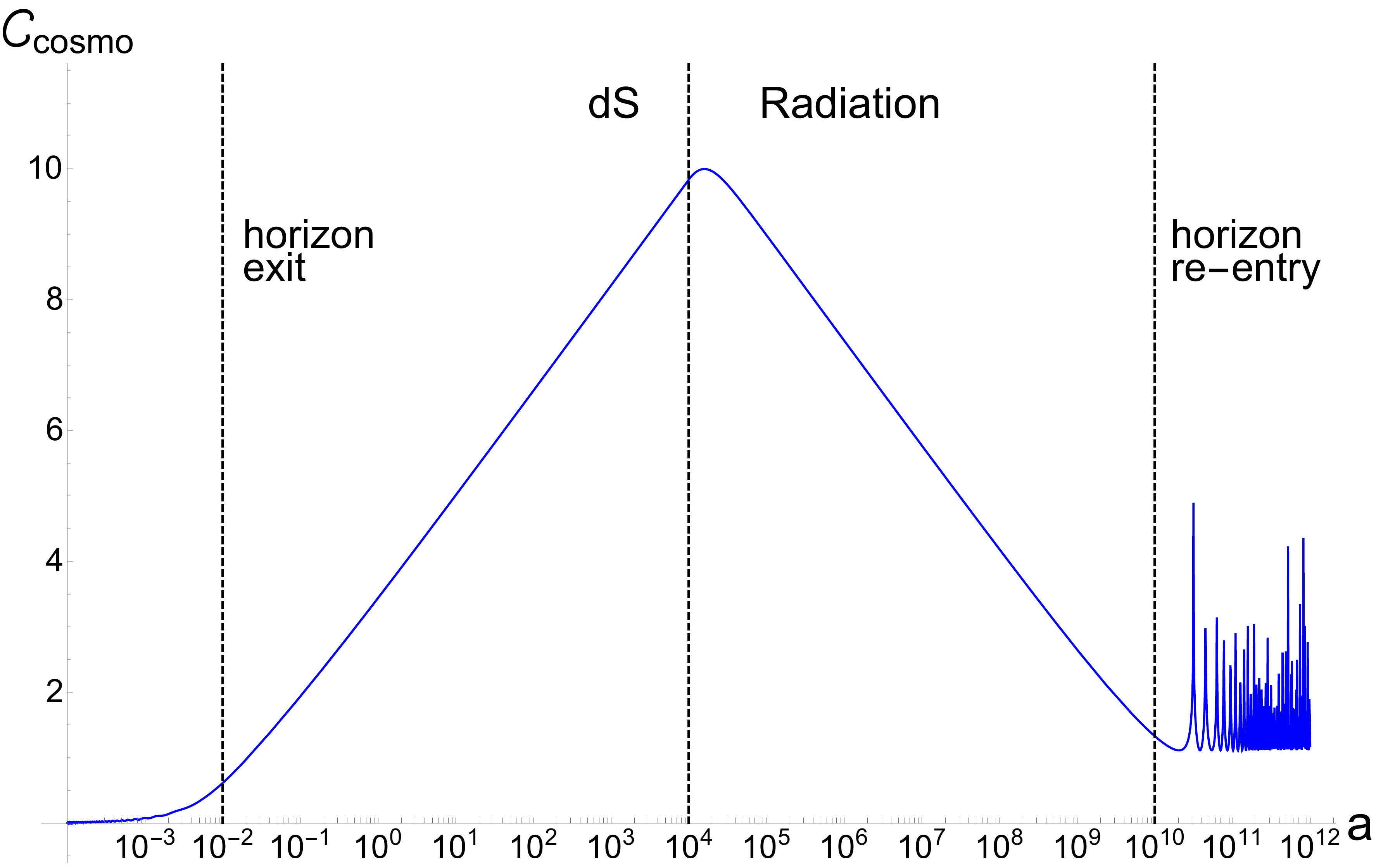}
	\caption{The complexity (\ref{eq:CosmoComplexity}) of a pure squeezed cosmological perturbation (\ref{psi1}) grows while the mode is outside the horizon during de Sitter accelerated expansion, but \emph{decreases} during radiation domination, leading to a period of \emph{de-complexification}. Once the mode re-enters the horizon, rapid oscillations in $e^{-2i\phi_k}$ lead to rapid oscillations in the complexity.}
	\label{fig:CosmoComplexity}
\end{figure}

The ground state defined by $\hat c_{\vec{k}} |0_{\vec{k}}\rangle = 0$ at early times evolves at late times into the cosmological squeezed state \cite{Grishchuk,Albrecht}
\be
|\Psi_{\rm cosmo}\rangle_{\vec{k},-\vec{k}} = \frac{1}{\cosh r_k} \sum_{n=0}^{\infty} (-1)^n e^{-2 i n \varphi_k} \tanh^n r_k\, |n_{\vec{k}},n_{-\vec{k}}\rangle\, .
\label{psi1}
\ee
The squeezing parameter $r_k$ and squeezing angle $\phi_k$ depend on time through the time-dependence of the scale factor \cite{Albrecht,cosmology1,cosmology2}
\be
\frac{dr_k}{da} &=& -\frac{1}{a} \cos(2\phi_k)\, , \\
\frac{d\phi_k}{da}&=& \frac{k}{a{\mathcal H}} + \frac{1}{a} \coth(2r_k) \sin(2\phi_k)\, .
\ee
The resulting complexity of this squeezed state relative to the ground state \cite{cosmology1,cosmology2}
\be
{\mathcal C}_{\rm cosmo} = \frac{1}{\sqrt{2}} \sqrt{\left|\ln \left|\frac{1+e^{-2i\phi_k}\ \tanh r_k}{1-e^{-2i\varphi_k}\ \tanh r_k}\right|\right|^2+\arctan\left(2\sin 2\varphi_k\, \sinh r_k\, \cosh r_k\right)^2}\, .
\label{eq:CosmoComplexity}
\ee
also depends on time through the time-dependent squeezing parameter and squeezing angle $r(\eta),\varphi(\eta)$.
For example, following \cite{cosmology1} we consider a simple model of the very early Universe in which the expanding background starts as de Sitter (inflation) with Hubble constant $H_{dS}$, 
then transitions to a radiation-dominated expansion.
For a fixed comoving wavelength $k$ starting inside the horizon at early times $k\gg aH_{dS}$, the squeezing parameter is small $r_k(a) \approx a H_{dS}/k \ll 1$ and the squeezing angle is constant $\varphi_k \approx -\pi/4$; the corresponding complexity (\ref{eq:CosmoComplexity}) is small, as the state (\ref{psi1}) is still very nearly identical to the ground state.
As the accelerating universe expands, the mode eventually exits the horizon.
After horizon exit, the squeezing parameter begins to grow as the log of the scale factor $r_k \approx \ln a$ and the squeezing angle shifts to $\varphi \approx -\pi/2$, up to subleading corrections. The corresponding complexity also grows as the log of the scale factor at this time ${\mathcal C}_{\rm cosmo} \approx \ln(a)/\sqrt{2}$.
As the Universe transitions to radiation-domination, the squeezing parameter continues to grow as the log of the scale factor, but the squeezing angle now is repelled away from $\varphi \approx -\pi/2$ towards positive values.
Curiously, even though the squeezing parameter is growing, the growing squeezing angle causes the complexity to \emph{decrease} during radiation domination\footnote{There is a slight delay to the onset of de-complexification after radiation domination as the slope of the squeezing angle gradually changes sign, as discussed in \cite{cosmology1}.}, leading to a period of \emph{de-complexification}.
At late times the mode re-enters the horizon and the squeezing parameter ``freezes in'' to a constant $r_k\approx r_*$ determined by horizon crossing. The squeezing angle when the mode re-enters the horizon becomes large and $k$-dependent, so that $e^{-2\varphi_k}$ oscillates rapidly. The resulting complexity also ``freezes in'' to a constant value, up to oscillations, at late times.
These behaviors for the squeezing parameter $r_k$, squeezing angle $\varphi_k$ (through $\cos \varphi_k$) are shown in Figure \ref{fig:CosmoSqueezing}, and the corresponding complexity ${\mathcal C}_{\rm cosmo}$ is shown in Figure \ref{fig:CosmoComplexity}.

The complexity associated to the state (\ref{psi1}) assumes that the cosmological perturbations remain in a pure state throughout their evolution on the classical expanding background, with density matrix
\be
\hat \rho_{\rm pure} = |\Psi_{sq}\rangle \langle \Psi_{sq}| = \frac{1}{\cosh^2r_k} \sum_{n,m = 0}^\infty (-1)^{n+m} e^{-2i(n-m)\varphi_k} \tanh^{n+m} r_k\, |n_{\vec{k}},n_{-\vec{k}}\rangle\langle m_{\vec{k}}, m_{-\vec{k}}|\, . \hspace{.2in}
\ee
In general, however, we expect the perturbations to experience decoherence at some point in their evolution, 
as the quantum excitations freeze-in as classical density perturbations of the cosmic fluid.
There are many interesting models and descriptions of this process of cosmological decoherence, see \cite{Brandenberger_1992,Brandenberger_1993,Burgess_2008,Martin_2018,Shandera_2018,Gong_2019} and references therein for some examples.
Motivated by the rapid $k$-dependent growth of the squeezing angle $\varphi_k$ when the mode re-enters the horizon during the radiation era, with corresponding rapid oscillation of $e^{-2\varphi_k}$ as seen in Figure \ref{fig:CosmoSqueezing}, a simple model for decoherence is to average the density matrix over the squeezing angle \cite{Brandenberger_1992,Brandenberger_1993,Brahma:2020zpk}. The resulting reduced density matrix has only diagonal entries
\be
\hat \rho_{\rm red} = \frac{1}{\cosh^2r_k} \sum_{n= 0}^\infty \tanh^{2n} r_k\, |n_{\vec{k}},n_{-\vec{k}}\rangle\langle n_{\vec{k}}, n_{-\vec{k}}|\, .
\label{eq:CosmoReducedRho}
\ee
This reduced density matrix (\ref{eq:CosmoReducedRho}) has the form of a thermal density matrix $\hat \rho_{\rm th}$ of a harmonic oscillator
with temperature $\beta = -\ln (\tanh^2 r_k)$.
For modes inside the horizon the squeezing parameter is approximately constant $r_k \approx r_*$, so the resulting temperature of modes in this mixed state is also constant, $\beta \approx -\ln(\tanh^2 r_*)$.
Note, however, that this temperature is different from the Gibbons-Hawking temperature of de Sitter space $T_{GH} \sim H/2\pi$ \cite{Gibbons:1977mu}, as it depends on the evolution history of the mode and the amount of growth it experiences when it is outside the horizon.

\begin{figure}[t]
	\centering \includegraphics[width=0.95\textwidth]{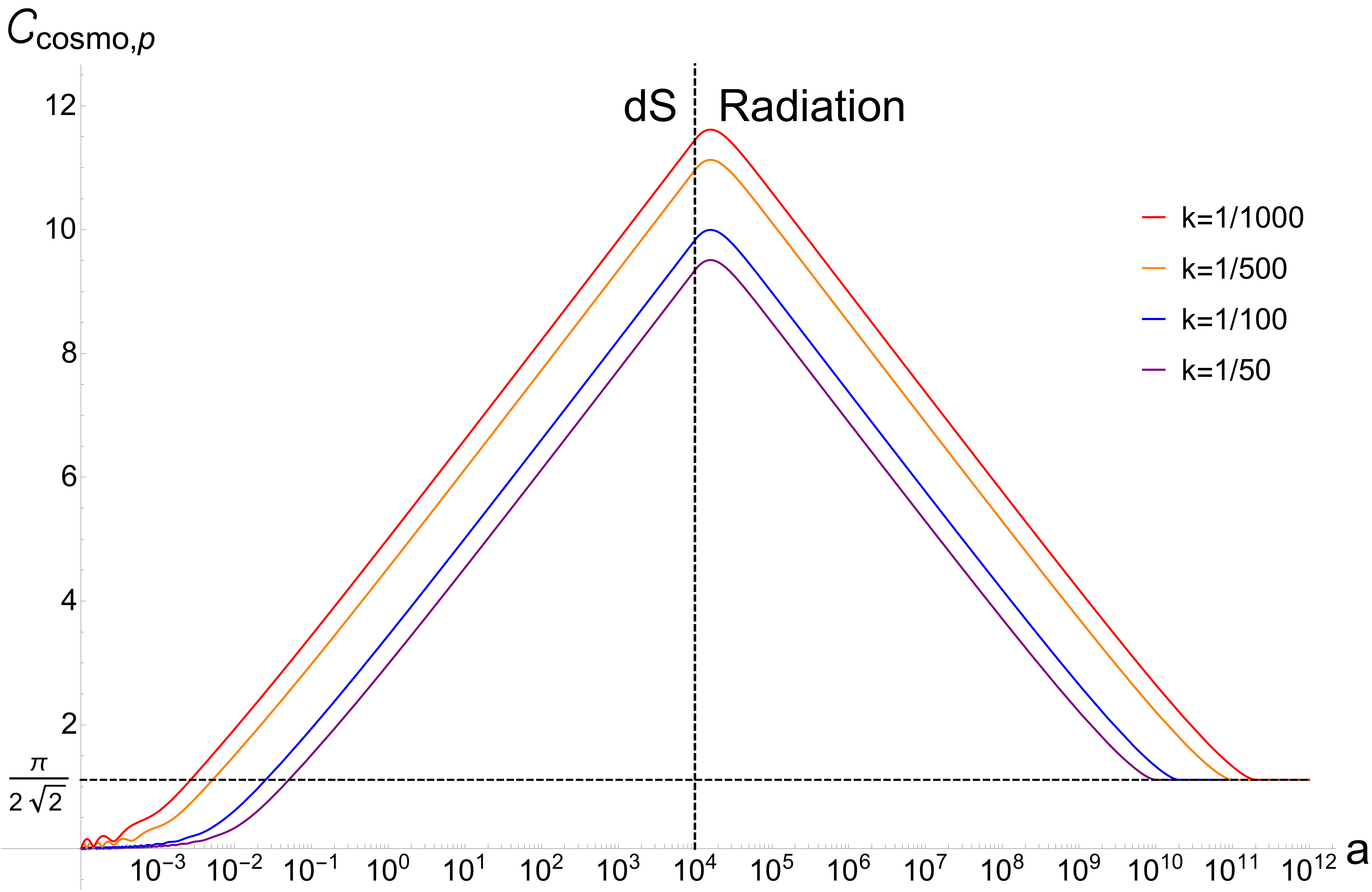}
	\caption{The complexity for cosmological perturbations in the simple cosmological background of Figures \ref{fig:CosmoSqueezing},\ref{fig:CosmoComplexity} with a simple model of decoherence in which the perturbations are described by the pure state (\ref{psi1}) until they re-enter the horizon in the radiation era, after which they are described by the mixed state (\ref{eq:CosmoReducedRho}) with corresponding complexity (\ref{eq:CosmoComplexPurif}). We see that all $k$-modes saturate to ${\mathcal C}_{\rm cosmo,p} \approx \pi/(2\sqrt{2})$ upon horizon re-entry.}
	\label{fig:CosmoPureComplexity}
\end{figure}

Since the reduced density matrix (\ref{eq:CosmoReducedRho}) resulting from our simple model of decoherence is thermal, we can calculate the associated thermal complexity of purification of the cosmological perturbations.
To do this, we expand our Hilbert space to include an ancillary copy (suppressing the $-\vec{k}$ mode index), ${\mathcal H_{\vec{k}}}\otimes {\mathcal H'_{\vec{k},{\rm anc}}}$, resulting in the purification
\be
|\Psi_{\rm cosmo,p}\rangle_{\vec{k}} = \frac{1}{\cosh r_k} \sum_{n=0}^{\infty} (-1)^n e^{-2 i n \phi} \tanh^n r_k\, |n_{\vec{k}}\rangle \otimes |n'_{\vec{k}}\rangle\, .
\label{eq:CosmoPurif}
\ee
The purification (\ref{eq:CosmoPurif}) and the original cosmological squeezed state (\ref{psi1}) look nearly identical. The primary difference is that in the original squeezed state (\ref{psi1}) the squeezing parameter and squeezing angle $\varphi_k$ are determined by the dynamics of the expanding background, while for the purification (\ref{eq:CosmoPurif}) the squeezing parameter is still dynamical, in principle, but the purification squeezing angle $\phi$ is fixed to $\phi = \pi/4$ by the minimization of the complexity.
This modest difference modifies the behavior of the cosmological complexity in qualitative ways.
In particular, the complexity of the purified state (\ref{eq:CosmoPurif}) for the minimized value $\phi = \pi/4$ becomes
\be
{\mathcal C}_{\rm cosmo,p}(k) = \frac{1}{\sqrt{2}} \arctan(\sinh(2r_k))\, .
\label{eq:CosmoComplexPurif}
\ee
As noted before, this complexity has universal behavior as a function of the squeezing parameter $r_k$. In particular, for large squeezing $r_k \gg 1$ the complexity saturates ${\mathcal C}_{\rm cosmo,p} \approx \pi/(2\sqrt{2})$.
More generally, since this complexity is independent of the dynamical squeezing angle $\varphi_k$ it can have qualitatively different behavior than the pure state complexity (\ref{eq:CosmoComplexity}) when the reduced density matrix (\ref{eq:CosmoReducedRho}) is an appropriate description.

\begin{figure}[t]
	\centering \includegraphics[width=.9\textwidth]{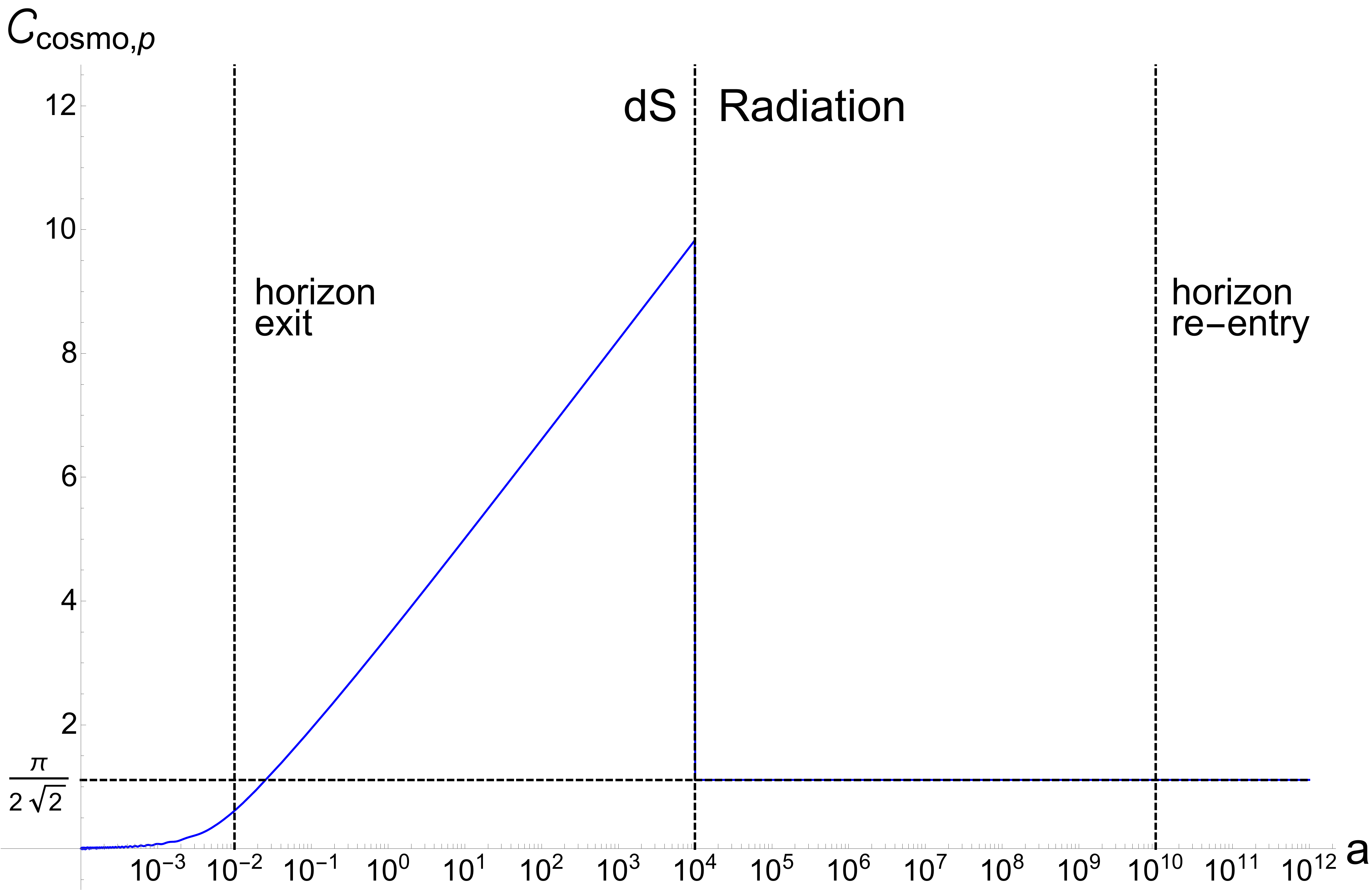}
	\caption{The complexity for cosmological perturbations with an alternative model of decoherence in which the perturbations evolve into the mixed state (\ref{eq:CosmoReducedRho}) at the transition between the de Sitter and radiation phases. The complexity jumps discontinuously to the saturated value ${\mathcal C}_{\rm cosmo,p} \approx \pi/(2\sqrt{2})$ at the moment of decoherence.}
	\label{fig:UniverseDecohere}
\end{figure}

Let's return to the simple model described earlier in which the Universe begins in a phase of accelerated de Sitter expansion, followed by a period of radiation domination.
A simple model of decoherence of the density perturbations in this background is to
take the density perturbation modes to be described by the pure state (\ref{psi1}) up until horizon re-entry in the radiation phase, after which the squeezing angle becomes large and $k$-dependent, triggering the averaging leading to the thermal density matrix (\ref{eq:CosmoReducedRho}).
The resulting complexity is shown in Figure \ref{fig:CosmoPureComplexity} for several different wavelengths $k$, and has the following qualitative features: while the mode is described by the pure state (\ref{psi1}), the complexity follows (\ref{eq:CosmoComplexity}) as seen in Figure \ref{fig:CosmoComplexity}. Since each $k$-mode has a slightly different horizon-exit time, the corresponding total growth in the complexity is different, while the slope is universal (see \cite{cosmology2} for more discussion of the slope of complexity for different cosmological backgrounds). 
Upon the transition to radiation-domination the complexity for all of the $k$-modes begins to decay.
As each mode re-enters the horizon, the corresponding squeezing angle begins to grow rapidly, triggering our model of decoherence (\ref{eq:CosmoReducedRho}),(\ref{eq:CosmoPurif}).
The corresponding complexity (\ref{eq:CosmoComplexPurif}) saturates to ${\mathcal C}_{\rm cosmo,p} \approx \pi/(2\sqrt{2})$ for each mode inside the horizon, smoothing out the rapidly oscillating complexity seen in Figure \ref{fig:CosmoComplexity}.
This simple model suggests that the complexity of each cosmological wavelength is universal and is equal to the saturated value after decoherence, once the mode re-enters the horizon.

While the above model of phase decoherence (or ``re-entry decoherence'') seems quite natural, let us briefly consider what the effects on complexity would be for decoherence of the form (\ref{eq:CosmoReducedRho}) occurring at different times.
If decoherence occurs at an earlier time than horizon re-entry, the complexity of the state will abruptly jump from its current value to the
constant ${\mathcal C}_{\rm cosmo,p} \approx \pi/(2\sqrt{2})$.
For example, if decoherence of inflationary density perturbations occurs at the transition between the de Sitter and radiation phases, the complexity for a given superhorizon mode will jump discontinuously, as in Figure \ref{fig:UniverseDecohere}.
Naturally, if the decoherence occurs at some other moment in the cosmological history we expect a corresponding discontinuous jump in the complexity.

\section{Discussion}
\label{sec:Discussion}

We constructed a purification of a thermal density matrix of a harmonic oscillator as a generic two-mode squeezed state,
where the squeezing is fixed by the temperature and the squeezing angle is a free parameter.
Minimizing the complexity relative to the ground state with respect to the squeezing angle, we found that the complexity \emph{saturates} at high temperatures $\beta \omega \ll 1$ to ${\mathcal C}_{\rm th} \rightarrow \frac{\pi}{2\sqrt{2}}\approx 1.1$ with a non-zero squeezing angle $\phi_{\rm min} = \pi/4$, in contrast to previous results \cite{MyersMixed,DiGiulio:2020hlz,Camargo:2020yfv}.
Since all purifications of the thermal density matrix result in identical expectation values, this implies a bound on the complexity of a thermal state ${\mathcal C}_{\rm th} \leq {\mathcal O}(1)$ for all temperatures and frequencies.
While this geometrized notion of complexity is not identical to a counting of the number of discrete unitary operators needed to build up the target state from the reference state, the bound indicates that the number of such operators needed to construct a minimal purification of a thermal state is similarly bounded.
We considered further purifications, including an additional squeezing of the ancillary degree of freedom, and demonstrated that the minimum complexity remains the same and is similarly bounded, although the additional squeezing allows more freedom and degeneracy in the minimized purification.
We further showed how operator-state mapping, which also associates a pure state to the thermal density matrix, does not minimize the complexity because it does not scan over all allowed squeezing angles. 
The simplest purification of a thermal state with a minimized complexity is thus a two-mode squeezed state with squeezing angle $\phi_{\rm min} = \pi/4$.

More generally, by analyzing the complexity of a two-mode squeezed state as a function of both the squeezing angle and squeezing parameter, 
we found that the complexity generically saturates for a non-zero squeezing angle. In contrast, for squeezing angles $\phi = n\pi/2$ the complexity is maximized and grows linearly with the squeezing parameter ${\mathcal C}_{\phi = n\pi/2} \approx r$ for large $r$.
Interestingly, several solutions of quantum fields on curved spacetimes can also be described as two-mode squeezed states with particular values of the squeezing angle.
For example, vacuum states of a scalar field on a curved background with a horizon in one basis, such as Rindler space or a black hole spacetime, can naturally be described as a two-mode squeezed vacuum state (``Unruh'' state) entangling modes on either side of the horizon.
The squeezing parameter is set by the inverse temperature of the horizon, while the squeezing angle is fixed at the maximal value $\phi = \pi/2$.
The complexity of these Unruh states, therefore, is also maximal and grows logarithmically with temperature ${\mathcal C}_{\rm Unruh} \sim \ln(T/\omega)$ at large temperature.
In a different context, cosmological curvature perturbations are also described as a two-mode squeezed state, entangling modes with opposite momenta.
For accelerating backgrounds, such as de Sitter, these modes are dynamically driven on superhorizon scales to high squeezing and squeezing angle $\phi\rightarrow -\pi/2$.
Correspondingly, the complexity for these modes continues to grow during the accelerating phase as the number of e-folds since horizon exit
${\mathcal C}_{\rm cosmo} \sim \ln(a(t)/a_{\rm exit})$.
Without the perspective of complexity, the dynamical preference of these backgrounds for the squeezing angle $\phi = n\pi/2$ would be opaque; note, for example, that the entanglement entropy of these squeezed states is independent of the squeezing angle.
Other quantities, such as Hawking radiation for backgrounds with horizons and the power spectrum for cosmological perturbations, are also independent of the squeezing angle.
However, complexity\footnote{Other measures of complexity of two-mode squeezed states, particularly those based on the correlation matrix \cite{DiGiulio:2020hlz}, appear to be independent of the squeezing angle.} 
appears to be unique in being sensitive to the squeezing angle. 
Moreover, the squeezing angle dynamically selected maximizes the complexity of the full state, suggesting that these backgrounds may be seeking to find a field configuration that maximizes the complexity in a form of a second law of complexity \cite{Brown:2017jil}.
It would be interesting to study this intriguing possibility further.

We applied our minimal purification to thermal density matrices that arise from these interesting curved space examples under certain conditions.
For example, tracing out the modes beyond the horizon, an observer in a Rindler or black hole spacetime sees a thermal distribution of Hawking radiation. 
While the global solution includes modes on both sides of the horizon and maximizes the complexity, 
an observer can purify the thermal density matrix of Hawking radiation with a smaller complexity by choosing a squeezing angle $\phi_{\rm min} = \pi/4$ instead.
This purified two mode squeezed state resembles the global solution, but on a space in which the observable modes are entangled with ancillary modes on a ``twisted'' wedge beyond the horizon.
The reduced complexity of the minimized purification seems to suggest that it is possible to model the radiation external to a black hole with fewer operators than the global solution would have suggested. It would be interesting to see if other information-theoretic quantities of the black hole depend sensitively on the squeezing angle, and thus might demonstrate some preference for some squeezing angles over others.

For cosmological perturbations, we implemented a simple model of decoherence where the squeezing angle is averaged over, leading to a thermal density matrix with a temperature determined by the squeezing of the mode.
Implementing this model of decoherence when modes re-enter the horizon during the radiation-dominated expansion of the Universe, 
the resulting complexity smoothly matches to the preceding period of de-complexification, saturating at ${\mathcal C}_{\rm cosmo} \approx \pi/(2\sqrt{2})$, see Figure \ref{fig:CosmoPureComplexity}.
Alternatively, implementing this model of decoherence at other times, such as at the transition between de Sitter- and radiation-domination, leads to a sharp drop in the purified complexity.
This suggests that complexity might be a sensitive measure of different models and implementations of decoherence.

\section*{Acknowledgements}

We would like to thank Arpan Bhattacharyya and Sayura Das for helpful conversations and discussions. S.H.~would like to thank the University of Cape Town for funding this project.

\appendix

\section{Complexity of Two-Mode Squeezing}
\label{app:Complexity}

In this section, we review the calculation of the quantum circuit complexity for two-mode squeezing, following \cite{NL1,Jefferson,me1}.
A reference state $|\psi_R\rangle$ is transformed into a target state $|\psi_T\rangle$
\be
|\psi_T\rangle = \hat {\mathcal U}\ |\psi_R\rangle
\label{eq:NielsenApp}
\ee
by a unitary operator $\hat {\mathcal U}$ representing the quantum circuit connecting these two states.
The unitary $\hat {\mathcal U}$ can be written as a path-ordered exponential constructed as a sequence of gates
\be
\hat U(s) = \overleftarrow{\mathcal P}\ {\rm exp}\left[-i \int_0^s \sum_I Y^I(s') \hat {\mathcal O}_I\  ds'\right]
\label{eq:UnitaryApp}
\ee
where $s$ parameterizes a path the circuit takes in the space of operators, $\hat {\mathcal O}_I$ are a set of operators that will act as our gates,
and the $Y^I(s)$ are vectors that specify the path in the space of operators.
The path is parameterized with the boundary conditions such that $s=0$ corresponds to the identity operator $\hat U(0) = \mathds{1}$ while $s=1$ corresponds to the final state $\hat U(1) = \hat {\mathcal U}$.

The \emph{quantum circuit complexity} (denoted as ``complexity'') for this construction is defined as the minimum circuit depth
\be
{\mathcal C} = \int_0^1 {\mathcal F}(\hat U)\ ds\, ,
\ee
where ${\mathcal F}(\hat U)$ is a \emph{cost function}, which must be chosen by hand.
The minimum complexity then amounts to finding the path $Y^I(s)$ that generates the unitary $\hat U(s)$ subject to the boundary conditions, and minimizes the cost function ${\mathcal F}(Y)$.
Several different choices of cost functions exist; some common choices are \cite{NL1,NL2,NL3,Jefferson}
\be
{\mathcal F}_1(Y) &=& \sum_I |Y^I|\, ,\hspace{.2in} {\mathcal F}_2 = \sqrt{\sum_I G_{IJ} Y^I Y^J}\, , \hspace{.2in} {\mathcal F}_{\kappa} = \sum_I |Y^I|^{\kappa}\, .
\ee
The ${\mathcal F}_1$ cost function is useful in that it directly counts the number of gates, and is clearly extensive, but the result depends on the basis chosen for the operators.
The ${\mathcal F}_2$ cost function, on the other hand, is clearly the total length of a geodesic, but is not extensive. The metric $G_{IJ}$ represents a cost or penalty factor for certain operators. We will primarily work with the ${\mathcal F}_2$ cost function, and will choose a flat Riemannian metric with no penalty factors $G_{IJ} = \delta _{IJ}$ for simplicity.

We consider two-mode states in which the Hilbert space ${\mathcal H}\otimes {\mathcal H}_{\rm anc}$ consists
of two copies of a quantum harmonic oscillator.
We will take as our reference state the ground state of a harmonic oscillator $|\psi_R\rangle = |0\rangle\otimes |0\rangle_{\rm anc}$ with (real) reference frequency $\omega_R$
\be
\langle q,q_{\rm anc}|\psi_R\rangle = {\mathcal N}_R\ {\rm exp}\left[-\frac{1}{2} \omega_R (q^2 + q_{\rm anc}^2)\right]\, .
\label{eq:gaussianRefApp}
\ee
Our target state is the two-mode squeezed state $|\psi_T\rangle = |\Psi\rangle_\phi$ (\ref{eq:TFDsqueeze}) 
\be
\Psi_{\rm sq}\left(q,q_{\rm anc}\right) = \langle q, q_{\rm anc}|\Psi\rangle_{\phi} = {\cal N}_T\ \exp\left\{-\frac{\omega}{2}\, A (q^2 + {q}_{\rm anc}^2) - \omega B\ q\ q_{\rm anc} \right\}
\label{eq:gaussianTargetApp}
\ee
where
\begin{equation}
	\begin{split}
		&
		A = \frac{1 + e^{-4i\phi} \tanh^2 r}{1-e^{-4i\phi} \tanh^2 r} \,,  
		\qquad
		B = \frac{2\tanh r\, e^{-2i\phi}}{1- e^{-4i\phi} \tanh^2 r} \,.
	\end{split}\label{eq:gaussianABApp}
\end{equation}
The wavefunctions (\ref{eq:gaussianRefApp}),(\ref{eq:gaussianTargetApp}) can be diagonalized in terms of rotated coordinates $\tilde{q}_1, \tilde{q}_2$
\be
{\rm exp}\left[-\frac{1}{2} \tilde{q}_a A_{ab} \tilde{q}_b\right], \hspace{.2in} \vec{\tilde q} = \begin{pmatrix}\tilde{q}_1 \cr \tilde{q}_{2}\end{pmatrix}
\ee
so that the symmetric square matrices representing the reference and target states take the form
\be
A_R = \omega_R \mathds 1, \hspace{.2in} A_T = \omega \begin{pmatrix}\Omega_1 & 0 \cr 0 & \Omega_2\end{pmatrix}\, 
\ee
where $\Omega_1 = -2A + B, \Omega_2 = -2A - B$.

The unitary operator (\ref{eq:UnitaryApp}) acts on the reference matrix as $A(s) = \hat U(s) A_R \hat U^T(s)$.
Since the entries (\ref{eq:gaussianABApp}) can be complex, we will use the diagonal elements of $GL(N,\mathbb{C})$ as our set of gate operators $\hat {\mathcal O}_I$ (\ref{eq:UnitaryApp})
\be
\hat U(s) = {\rm exp}\left[y^a(s) M_a^{\rm diag} \right] = {\rm exp}\left[\begin{pmatrix}y^1(s) & 0 \cr 0 & y^2(s)\end{pmatrix}\right]
\ee
where the $y^a(s) = \alpha^a(s) + i \beta^a(s)$ are complex. The resulting metric on the reduced space of operators 
becomes \cite{me1}
\be
ds^2 = G_{IJ} dY^I dY^J = \sum_{a=1}^2 |y^a|^2 = \sum_{a=1}^2 \left[(\alpha^a)^2 + (\beta^a)^2\right]\, .
\ee
The resulting complexity
\be
{\mathcal C} = \int_0^1 \sqrt{G_{IJ} dY^I dY^J}\ ds = \int_0^1 \sqrt{\sum_{a=1}^2 \left[(\alpha^a)^2 + (\beta^a)^2\right]}\ ds,
\ee
is minimized, subject to the boundary conditions, by the straight-line geodesic
\be
\alpha^a(s) &=& \ln \left|\Omega_a\frac{\omega}{\omega_R}\right|\ s, \hspace{.3in} \beta^a(s) = \arctan \left[\frac{{\rm Im}(\Omega_a)}{{\rm Re}(\Omega_a)}\right]\ s\ .
\ee
Choosing the reference frequency to be the ground state frequency of the oscillator $\omega _R = \omega$ leads to the minimized complexity used in the main text 
\be
{\mathcal C} = \frac{1}{2}\sqrt{\sum_{a=1}^2 \left[\left(\ln |\Omega_a|\right)^2 + \left(\arctan \left[\frac{{\rm Im}(\Omega_a)}{{\rm Re}(\Omega_a)}\right]\right)^2\right]}\, .
\ee

\section{Bogoliubov Transformations and Squeezed States}
\label{app:Bogo}

We start with a massless scalar field expanded in modes
\be
\hat \phi = \sum_i \left(u_i \hat a_i + u_i^* \hat a_i^\dagger\right)
\ee
where $\hat a_i,\hat a_i^\dagger$ are creation and annihilation operators in the basis $(u_i, u_i^*)$.
The vacuum state for this basis is defined by $\hat a_i |0\rangle_a = 0$ for all $\hat a_i$.
In curved spacetimes or quench models, it is often possible to expand $\hat \phi$ in another basis
\be
\hat \phi = \sum_j \left(v_j \hat b_j + v_j^* \hat b_j^\dagger \right)\, ,
\ee
with its own distinct vacuum state defined by $\hat b_j |0\rangle_b = 0$.
Since both sets are complete, we can expand the annihilation operators in the original basis in terms of the new basis
through a Bogoluibov
transformation
\be
\hat a_i = \sum_{j} \left[\alpha_{ij} \hat b_{j} + \beta^*_{ij} \hat b_{j}^\dagger\right]\, ,
\label{eq:BogoGeneral}
\ee
where the Bogoluibov coefficients must satisfy the conditions
\be
\sum_{k} \left(\alpha_{ik} \alpha^*_{jk} - \beta_{ik}\beta^*_{jk}\right) = \delta_{ij}\, .
\label{eq:BogoConsistency}
\ee

We are primarily interested in mixings between only two modes, which we will call ${\rm R}$ and ${\rm L}$, in the following 
way\footnote{In the examples discussed in the main text, spacetime naturally splits into wedges which we will call ${\rm R}$ and ${\rm L}$, which is the motivation for this choice.}
\be
\label{eq:BogoKK1}
\hat a_{\rm R} &=& \alpha\, \hat b_{\rm R} + \beta^*\, \hat b^\dagger_{\rm L}\, ; \\
\hat a_{\rm L} &=& \alpha\, \hat b_{\rm L} + \beta^*\, \hat b^\dagger_{\rm R}\, .
\label{eq:BogoKK2}
\ee
The vacuum state of the $\hat b$ basis is populated with particles when viewed from the $\hat a$ basis, since 
\be
\hat a_{\rm R} |0\rangle_b = \beta^* \hat b^\dagger_{L} |0\rangle_b \neq 0
\ee
so that
\be
{}_b\langle 0|\hat a^\dagger_{\rm R} \hat a_{\rm R} |0\rangle_b = |\beta|^2\, .
\ee
The normalization condition (\ref{eq:BogoConsistency}) becomes $|\alpha|^2 - |\beta|^2 = 1$, so a natural parameterization
is
\be
\alpha = e^{-i\theta}\cosh r, \hspace{.2in} \beta = -e^{-i(\theta+2\phi)} \sinh r\, .
\label{eq:BogoSqueezeParam}
\ee

This choice of the parameterization of the Bogoluibov coefficients allows us to easily make contact with the squeezed state language, since
(\ref{eq:BogoKK1}),(\ref{eq:BogoKK2}) with (\ref{eq:BogoSqueezeParam}) can also be written as the transformation
\be
\hat a_{\rm R} = \hat{\mathcal U} ^\dagger\, \hat b_{\rm R}\, \hat{\mathcal U}\, ,
\ee
where $\hat {\mathcal U}$ 
\be
\hat {\mathcal U} = \hat {\mathcal R}(\theta)\, \hat {\mathcal S}(r,\phi) 
\ee
is constructed from combination of the two-mode squeeze and rotation operators,
\be
\hat {\mathcal S}(r, \phi) &=& \exp\left[\frac{r}{2} \left(e^{-2i\phi} \hat a_{\rm R}\hat a_{\rm L} - e^{-2i\phi} \hat a_{\rm R}^\dagger \hat a_{\rm L}^\dagger\right)\right]\, ; \\
\hat {\mathcal R}(\theta) &=& \exp\left[-i\theta \left(\hat a_{\rm R}^\dagger \hat a_{\rm R} + \hat a_{\rm L}^\dagger \hat a_{\rm L}\right)\right]\, .
\ee
Applying (\ref{eq:BogoKK1}) to the vacuum state $|0\rangle_a$ 
\be
\left(\alpha\, \hat b_{\rm R} + \beta^*\, \hat b^\dagger_{\rm L}\right) |0\rangle_a = 0
\ee
leads to the two-mode squeezed state solution
\be
|0\rangle_a = {\mathcal N} \exp\left[-\left(\frac{\beta^*}{\alpha}\right)\hat b_{\rm R}^\dagger \hat b_{\rm L}^\dagger\right] |0\rangle_b = {\mathcal N} \sum_{n=0}^\infty (-1)^n \left(\frac{\beta^*}{\alpha}\right)^n |n\rangle_{\rm R} \otimes|n\rangle_{\rm L}\, ,
\label{eq:BogoSqueezedState}
\ee
so that an alternative way of viewing the Bogoluibov transformations (\ref{eq:BogoKK1}),(\ref{eq:BogoKK2}) is that the vacuum state in one basis is a squeezed vacuum state (with non-vanishing particle number) in another basis. We note in passing that the relations (\ref{eq:BogoSqueezeParam}) allow us to extract the squeezing parameter $r$, and squeeze and rotation angles $\phi, \theta$ from the coefficients $\alpha, \beta$.
A thermal density matrix $\hat \rho_{\rm th}$ naturally arises from (\ref{eq:BogoSqueezedState}) upon tracing out over the ${\rm L}$-modes of the $\hat b$ basis:
\be
\hat \rho &=& {\rm Tr}_{\rm L}\left[|0\rangle_a\langle 0|_a\right] = {\rm Tr}_{\rm L}\left[|{\mathcal N}|^2 \sum_{mn} (-1)^{m+n} \left(\frac{\beta}{\alpha^*}\right)^{n}\left(\frac{\beta^*}{\alpha}\right)^{m} |n\rangle_{\rm R}\langle m|_{\rm R}\,\otimes |n\rangle_{\rm L}\langle m|_{\rm L}\right] \\
&=& |{\mathcal N}|^2 \sum_n \left|\frac{\beta}{\alpha^*}\right|^{2n}|n\rangle_{\rm R}\langle n|_{\rm R}\, .
\ee


   

\bibliographystyle{utphysmodb}

\bibliography{refs}

\end{document}